\documentclass[aip,reprint]{revtex4-1}

\usepackage{color}
\usepackage{graphicx}
\usepackage{amsmath}
\usepackage{subcaption}

\usepackage{natbib}

\begin{document}
\title{Kinetic Corrections to Heat-flow and Nernst advection for Laser Heated Plasmas}

\author{C. A. Walsh}
\email{walsh34@llnl.gov}
\affiliation{Lawrence Livermore National Laboratory}
\author{M. Sherlock}
\affiliation{Lawrence Livermore National Laboratory}

\date{\today}

\begin{abstract}
	Reduced models for approximating the impact of kinetic electron behavior on the transport of thermal energy and magnetic field are investigated. The thermal flux limiter has improved agreement with Vlasov-Fokker-Planck data when a harmonic form is used that adjusts the electron mean free path to account for electron-electron collisions; these results apply to both unmagnetized and magnetized plasmas. Once a magnetic field is incorporated, the mean free path should also be modified using the electron gyroradius. A flux limiter on Nernst advection of magnetic field is also required; a form that limits Nernst by the same fraction as the thermal heat-flow best reproduces kinetic simulations. A flux limiter form for the cross terms (Righi-Leduc and cross-gradient-Nernst) are also suggested. Hohlraum simulations relevant to fusion experiments on the National Ignition Facility are found to be sensitive to all of these details.
\end{abstract}
\maketitle

Codes used to design high energy density plasma experiments typically assume a classical plasma, where the distribution of electron velocities is close to Maxwellian. It has long been known that these plasmas tend to be kinetic in nature when they are heated with lasers; strong temperature gradients are formed, resulting in the distance between electron collisions being comparable to the temperature scale length. Non-local models, such as Vlasov-Fokker-Planck (VFP), more accurately represent behavior in this regime; while limited VFP studies have been carried out \cite{joglekar2016,hill2017,sherlock2020,hill2021} they are not yet widely available due to computational constraints. Therefore, reduced models that approximate kinetic behavior are incorporated into classical codes. The most common example is the heat-flow flux limiter\cite{malone1975,shvarts1981}, which reduces the classical unmagnetized Spitzer heat-flow \cite{spitzer1953} in radiation-hydrodynamics codes. 

Magnetic fields are spontaneously generated in high energy density plasmas; these fields are expected to magnetize the electron heat-flow in inertial confinement fusion (ICF) experiments, such as at the hot-spot edge \cite{walsh2017,walsh2023} and in hohlraums \cite{farmer2017}. Magnetic fields are also intentionally applied to ICF implosions \cite{moody2021a,hohenberger2012}, with enhanced performance observed in both directly and indirectly driven experiments \cite{moody2022,chang2011}. In both of these scenarios an extended-magnetohydrodynamics \cite{braginskii1965} framework is utilized to explain observed phenomena (such as enhancements in asymmetry when magnetic fields are applied to direct-drive implosions \cite{bose2021,walsh2020a}) or to make predictions (such as how the performance of pre-magnetized implosions scales with applied field strength \cite{hansen2020,sio2023,walsh2022}). 

This paper first revisits flux limitation of unmagnetized heat-flow and uses kinetic simulations to inform a modified approach. This new form is then extended to magnetized plasmas, where there are 3 separate heat-flow components. A flux limiter for magnetic transport down temperature gradients (the Nernst effect) is also proposed.

Non-locality can be most simply quantified by using the mean free path between electron-ion collisions \cite{bissell2015}:
\begin{equation}
	\lambda_{ei} = \frac{16 \pi T_e^2 \epsilon_0^2 }{n_i Z^2 e^4 log\Lambda_{ei}} \label{eq:lam_ei}
\end{equation}
where $T_e$ is the electron temperature, $\epsilon_0$ is the permittivity of free space, $n_i$ is the ion number density, $Z$ is the atomic number and $log\Lambda_{ei}$ is the Coulomb logarithm. If the distance between collisions is comparable to the length-scale of the system, the distribution of electron velocities becomes perturbed far from Maxwellian, modifying the electron transport.

Alternatively, a non-locality length-scale has been suggested in place of the electron-ion mean free path in order to obtain the correct dependence on atomic number \cite{davies2023}: 

\begin{equation}
\lambda_{d} = \lambda_{ei} \sqrt{\frac{Z^2 + 0.24Z}{Z+4.2}} \label{eq:lam_d}
\end{equation}
Where the Z-dependent factor approximately accounts for the role of
electron-electron collisions. Flux limiters throughout this paper are stated as functions of $\lambda_{ei}/L_T$ or $\lambda_{d}/L_T$, where $L_T = T_e / |\nabla T_e|$ is the temperature length-scale. Kinetic simulations have consistently found the temperature length-scale to be more important than the density length-scale for perturbing the underlying distribution of electron velocities \cite{sherlock2020,brodrick2018,davies2023}. This paper does not consider reduced models for kinetic ions.

The reduced models have been implemented into the extended-magnetohydrodynamics code Gorgon \cite{ciardi2007,chittenden2009,walsh2020}. Updated magnetized transport coefficients\cite{sadler2021,davies2021} are used, which give substantially improved results\cite{2021} when compared with Epperlein \& Haines \cite{epperlein1986}. Further information about the physics implementation will be given throughout this paper.

To explore the general behavior of flux inhibition and relate it to
more realistic flux-limiters, we use the K2 Vlasov-Fokker-Planck (VFP)
code, which is based on the KALOS approach to VFP simulation \cite{bell2006,sherlock2017} including the effect of a magnetic field. For the simulations reported
here, K2 represents the electron distribution function $f\left(x,\boldsymbol{v},t\right)$
as an expansion in spherical harmonics up to first order, $f\left(\boldsymbol{v}\right)\approx f_{0}^{0}\left(x,v,t\right)Y_{0}^{0}+f_{1}^{0}\left(x,v,t\right)Y_{1}^{0}+f_{1}^{1}\left(x,v,t\right)Y_{1}^{1}$,
where the $Y_{n}^{m}=Y_{n}^{m}\left(\theta,\varphi\right)$ are the
spherical harmonics and the $f_{n}^{m}$ are the expansion coefficients
\cite{bell2006}. Since the higher order terms are damped out by collisions,
this allows for an efficient computational model. The electric field
is obtained by solving the Ampere-Maxwell law and a background magnetic
field is directed perpendicular to the temperature gradient. The electron-electron
collision operator includes both the full isotropic Fokker-Planck
operator acting on $f_{0}^{0}$ as well as the anisotropic collision
operator acting on $f_{1}^{m}$, which is particularly important for
low-Z plasmas. These collision operators are given explicitly in \cite{sherlock2017}
and discussed more fully in e.g. \cite{tzoufras2011}. More details on
the simulation setup are given in the next section.

This paper is organized as follows. Section \ref{sec:heatflow_noB} considers heat-flow in an unmagnetized plasma, finding that a harmonic flux limiter more closely represents kinetic data than the hard cutoff that is typically used in radiation hydrodynamics codes. It is also suggested that electron-electron collisions should be incorporated into the length-scale. Section \ref{sec:heatflow} then extends this approach to a magnetized heat-flow, finding that the collision length-scale should also incorporate the Larmor radius of the electrons. Section \ref{sec:nernst} then extends this approach to the Nernst effect, while section \ref{sec:Xterms} proposes a flux limiter for both the Righi-Leduc heat-flow and cross-gradient-Nernst advection of magnetic fields. The impact of these changes are then demonstrated using hohlraum simulations in sections \ref{sec:NoMHD_impact} and \ref{sec:MHD_impact}, with the former ignoring magnetic fields and the latter incorporating self-generated magnetic fields .

\section{Unmagnetized Heat-flow Flux Limiter \label{sec:heatflow_noB}}

\begin{figure}
	\centering
	\centering
	\includegraphics[scale=0.5]{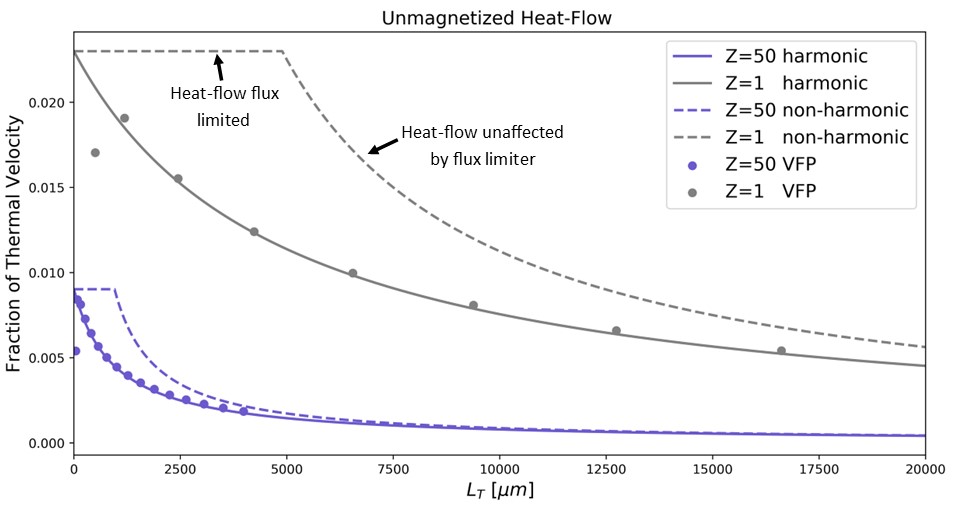}\caption{ \label{fig:FLM_unmag} Unmagnetized heat-flow normalized to the thermal velocity energy flux ($|\underline{q}|/n_e T_e v_{th}$) as a function of temperature length-scale.  Z=1 fits best with a flux limiter value of $f_{new} = 0.0035$ and Z=50 uses $f_{new} = 0.0025$. Most hydrodynamics codes use the non-harmonic flux limiter, despite the better representation of the data using a harmonic flux limiter.} 	
\end{figure}

First the unmagnetized heat-flow flux limiter will be revisited. Equivalently, this section applies to heat-flow along magnetic field lines in a magnetized plasma. For this reason, the unmagnetized thermal conductivity is written as $\kappa_{\parallel}$.

Thermal flux limiters are prevalent in radiation-hydrodynamics modeling of laser-plasma interactions, restricting the heat-flux velocity ($\underline{q}/n_e T_e$) to be a fraction ($f_{FL}$) of the electron thermal velocity ($v_{th}=(T_e/m_e)^{1/2}$) \cite{malone1975,shvarts1981}. Equivalently, this can be written in a form that shows the dependence on the degree of non-locality in the plasma:

\begin{equation}
\kappa^c _\parallel  \le f_{FL} \frac{L_{T\parallel}}{\lambda_{ei}} \label{eq:FL}
\end{equation}

The value of $f_{FL}$ is found empirically, typically taking a value in the range (0.05, 0.25). $\kappa^c _\parallel$ is the unmagnetized non-dimensional thermal conductivity that only depends on plasma ionization ($Z$) \cite{epperlein1986}. The dimensional thermal conductivity is related to the non-dimensional conductivity using $\kappa_{\parallel} = \kappa_{\parallel}^c n_e T_e \tau_{ei}/m_e$. Note that $L_{T\parallel}$ is the temperature length-scale along the magnetic field, which is equal to $T_e/\nabla T_e$ when no field is present. Considerations around calculating $L_{T\parallel}$ with a magnetic field are discussed in section \ref{sec:Code}.

The orthodox implementation of the thermal flux limiter is as a hard cut-off, which will be called the non-harmonic flux limiter:

\begin{equation}
\underline{q}_{\parallel nonharm} = -\frac{\kappa_\parallel \nabla T_e}{ max\Big(1,\frac{\kappa^c _\parallel \lambda_{ei}}{f_{FL} L_{T\parallel}}\Big)} \label{eq:q_nonharmonic}
\end{equation}

It is also possible to use a harmonic mean, which results in a smoother transition of heat-flow into the non-local regime:

\begin{equation}
\underline{q}_{\parallel harmonic} = -\frac{\kappa_{\parallel} \nabla T_e}{1 + \frac{\kappa_{\parallel}^c \lambda_{ei}}{f_{FL} L_{T\parallel}}} \label{eq:q_harmonic}
\end{equation}

1-D VFP simulations have been run in order to discern between the harmonic and non-harmonic flux limiter forms. The simulations consist of a single-sinusoid perturbation in the electron
temperature with small amplitude: $T_{e}\left(x\right)=T_{0}\left(1+0.02\sin kx\right)$,
with $k=2\pi/L$ and $T_0 = 500eV$. The simulation box size $L$ is adjusted to study
a variety of nonlocality parameters $\lambda_{ei}/L_{T}$, corresponding
to heat flows of different strengths. The electron density is uniform and 1.5$\times$10$^{26}$m$^{-3}$ in all cases. The applied
magnetic field ($B_{z}$) is also uniform but is varied to assess the scaling of non-local heat-flow suppression with magnetization. Cases for Z=1 and Z=50 are considered; these have mean free paths of $\lambda_{ei}=35\mu m$ and $0.7\mu m$ respectively. The initial temperature
perturbation induces a heat flow and electric field which can be recorded
in equilibrium and compared to their Braginskii counterparts. In order
to prevent the perturbation from relaxing too quickly, weak heating
and cooling operators are applied to $f_{0}^{0}$ which serve to maintain
a temperature profile close to the initial functional form (as used
in \cite{henchen2018,milder2022}). These simulations are similar in nature
to those originally carried out in \cite{epperlein1991} except
that (a) we also account for the magnetic field, (b) we monitor a
wider range of transport processes and (c) we include anisotropic
electron-electron collisions for improved accuracy. It is important
to note that the results obtained from these simulations cannot be
used in place of a VFP simulation for any general case. However, they
do provide insights into the sensible behavior and scaling of the
transport processes at different non-locality, and they offer enough
insight to design flux limiters that capture some of the general behavior,
as well as inform us how different processes (e.g. heat flow and the
Nernst effect) may have closely related scaling. This latter point
allows us to choose more consistent sets of flux limiters.

The VFP data in figure \ref{fig:FLM_unmag} shows the heat-flow increasing as the temperature length-scale decreases until a certain point at low $L_{T\parallel}$. 

Theoretical curves for the non-harmonic flux limiters are included as dashed lines (with flux limiter values of $f_{FL}=0.023$ for $Z=1$ and $f_{FL}=0.0045$ for $Z=50$). The non-harmonic flux limiter is always either on or off; when the limiter is on ($\kappa^c  > f_{FL} \frac{L_{T\parallel}}{\lambda_{ei}}$) the curves are flat (invariant to temperature length-scale) and when the limiters are off the heat-flow is classical.

The harmonic version of the flux limiter is plotted in figure \ref{fig:FLM_unmag} using solid lines. The gradual impact of the flux limiter on the heat-flow is better represented in this form, although the decrease in heat-flow at very low temperature length-scales is not reproduced. A flux limiter form that better represents the short temperature length-scale effects is investigated at the end of section \ref{sec:heatflow}.

Now focus turns to the dependence of the flux limiter value $f_{FL}$ on plasma ionization $Z$. The current orthodoxy in the field is to have a fixed flux limiter value within a simulation. However, the kinetic simulation data in figure \ref{fig:FLM_unmag} suggests a factor of 5 difference in flux limiter values between Z=1 and Z=50. Instead of using the electron mean free path ($\lambda_{ei}$) as the characteristic length-scale between collisions, the non-locality length-scale $\lambda_{d}$ (which incorporates electron-electron collisions) is now considered. Epperlein \& Short were the first to suggest this approach.

Epperlein \& Short conducted VFP simulations in order to find a reduced model for the thermal flux limiter in an unmagnetized environment \cite{epperlein1991}, much like the approach taken in this paper. The form they suggested can also be written as a harmonic flux limiter \cite{epperlein1991,davies2023}:

\begin{equation}
\underline{q}_{\parallel harmonic} = -\frac{\kappa_\parallel \nabla T_e}{1 + \frac{ \lambda_{d}}{f_{new} L_{T\parallel}}} \label{eq:q_harmonic_lambdad1}
\end{equation}

Importantly, this uses the non-locality length-scale $\lambda_{d}$ in place of $\lambda_{ei} \kappa ^c$, which gives a different dependence on plasma ionization. 

Note that Epperlein \& Short suggested a fixed value for the flux limiter $f_{new}$. In reality flux limitation is simply an approximation of more detailed kinetic behavior and the value of $f_{new}$ should be picked to best represent experiment data.

The harmonic flux limiter results in figure \ref{fig:FLM_unmag} can be obtained either by using the mean free path form (equation \ref{eq:q_harmonic}) or the non-locality length-scale (equation \ref{eq:q_harmonic_lambdad1}), just with different flux limiter values. When using the traditional mean free path form, the flux limiter values are $f_{FL}=0.023$ for $Z=1$ and $f_{FL}=0.0045$ for $Z=50$. When using the non-locality length-scale, the flux limiter takes the values $f_{new} = 0.0035$ for $Z=1$ and $f_{new} = 0.0025$ for $Z=50$. Using the non-locality length-scale results in less variation in the flux limiter with $Z$, and so is adopted here. This means a simulation can utilize a single flux limiter value, rather than having to tune experimental data against multiple free variables.

Further evidence for the use of the non-locality length-scale comes from the decrease in heat-flow at low temperature length-scales (as seen in figure \ref{fig:FLM_unmag}). The peak heat-flow occurs at different values of Knudsen number, if the traditional definition is used ($Kn = \lambda_{ei}/L_{T\parallel}$). If, instead, the non-locality length-scale is used in the definition, the peaks occur at the same value of adjusted Knudsen number ($Kn_d = \lambda_{d}/L_{T\parallel}$).

\begin{figure}
	\centering
	\centering
	\includegraphics[scale=0.5]{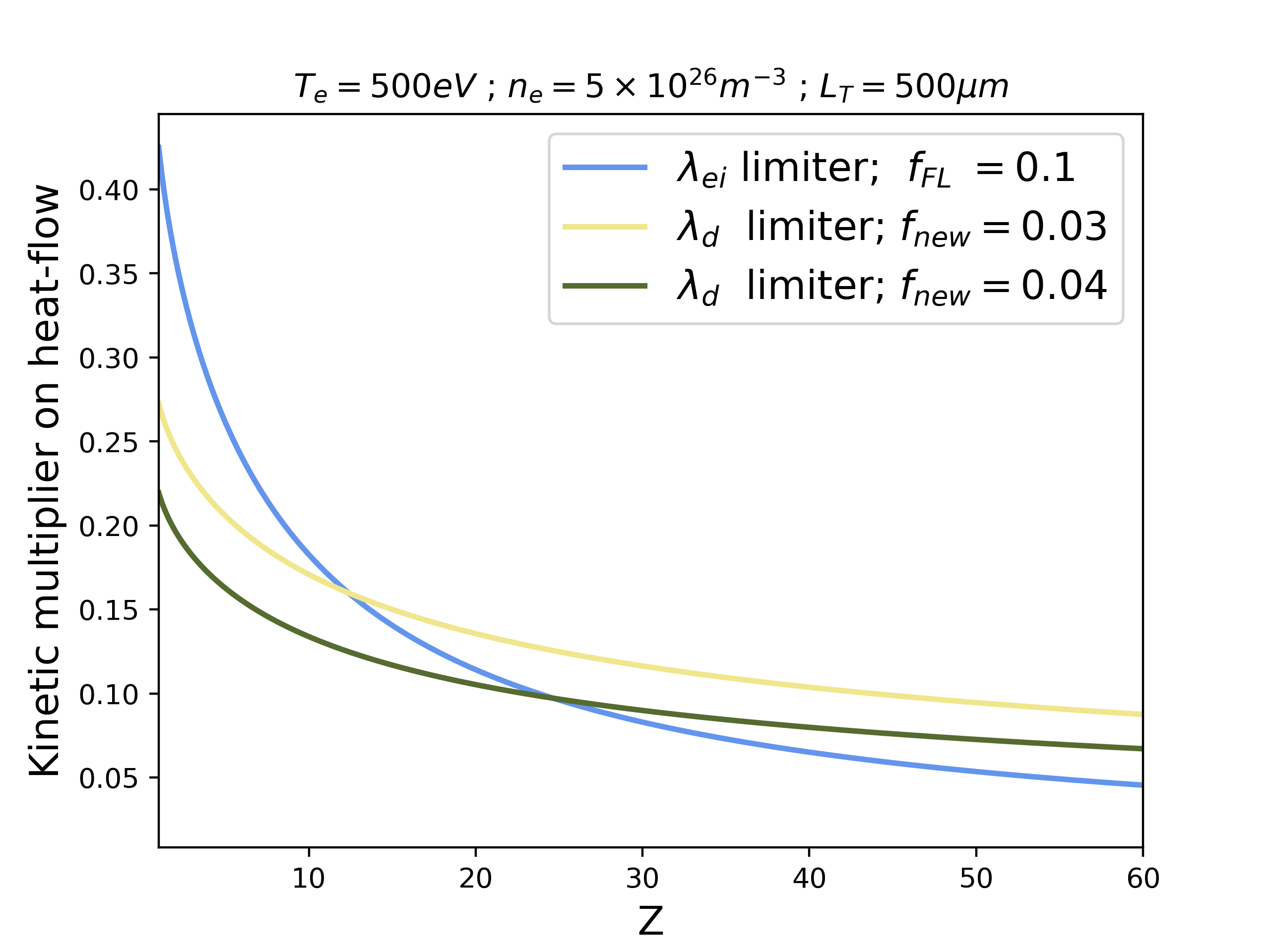}\caption{ \label{fig:FLM_vs_Z} Fractional restriction in unmagnetized heat-flow against plasma ionization for different forms of the flux-limiter, all using a harmonic form. The non-locality length-scale limiter ($\lambda_d$) is found in this paper to have better agreement with kinetic electron simulations, suggesting more restricted heat-flow at low Z and less restricted heat-flow at high Z. } 	
\end{figure}

Figure \ref{fig:FLM_vs_Z} shows the fractional suppression of unmagnetized heat-flow when using either the traditional mean-free path flux limiter or the non-locality limiter. Plasma parameters are $T_e=500eV$, $n_e = 5\times 10^{26} m^{-3}$, $L_{T\parallel} = 500\mu m$. The choice of flux limiter value $f_{new}$ can match the old flux limitation for a given plasma ionization, but the new flux limiter form generally suppresses heat-flow more for low Z and less for high Z.

\section{Magnetized Heat-flow Flux Limiter \label{sec:heatflow}}

When a magnetic field is applied to a plasma, the heat-flow becomes anisotropic \cite{braginskii1965}:

\begin{equation}
\underline{q}_e = -\kappa_{\parallel} \nabla_{\parallel}T_e - -\kappa_{\bot} \nabla_{\bot}T_e - \kappa_{\wedge} \underline{\hat{b}} \times \nabla T_e  \label{eq:magheatflow}
\end{equation}
Where the first term is the heat-flow along magnetic field lines (unchanged by magnetization), the second term is the  heat-flow perpendicular to field lines (reduced by magnetization), and the final term is the Righi-Leduc heat-flow (maximum at moderate magnetizations). This section deals with the perpendicular heat-flow by building on the unmagnetized results from section \ref{sec:heatflow_noB}. Again, the results are found to more closely follow a harmonic dependence rather than a hard cut-off. 

To account for magnetization in the heat-flow perpendicular to the magnetic field, the following form is used for the flux limiter:

\begin{equation}
\underline{q}_{\bot harmonic} = -\frac{\kappa_\bot \nabla_\bot T_e}{1 + \frac{\lambda_{d}}{f_{new} L_{T\bot}}\frac{\kappa_\bot^c}{\kappa_\parallel^c}} \label{eq:q_harmonic_lambdad}
\end{equation}

This form is also tested against VFP data. In the kinetic simulations the magnetic field is held uniform, otherwise the Nernst term acts to modify the magnetic field strength. 

\begin{figure}
	\centering
	\begin{subfigure}{\linewidth}
		\centering
		\includegraphics[scale=0.3]{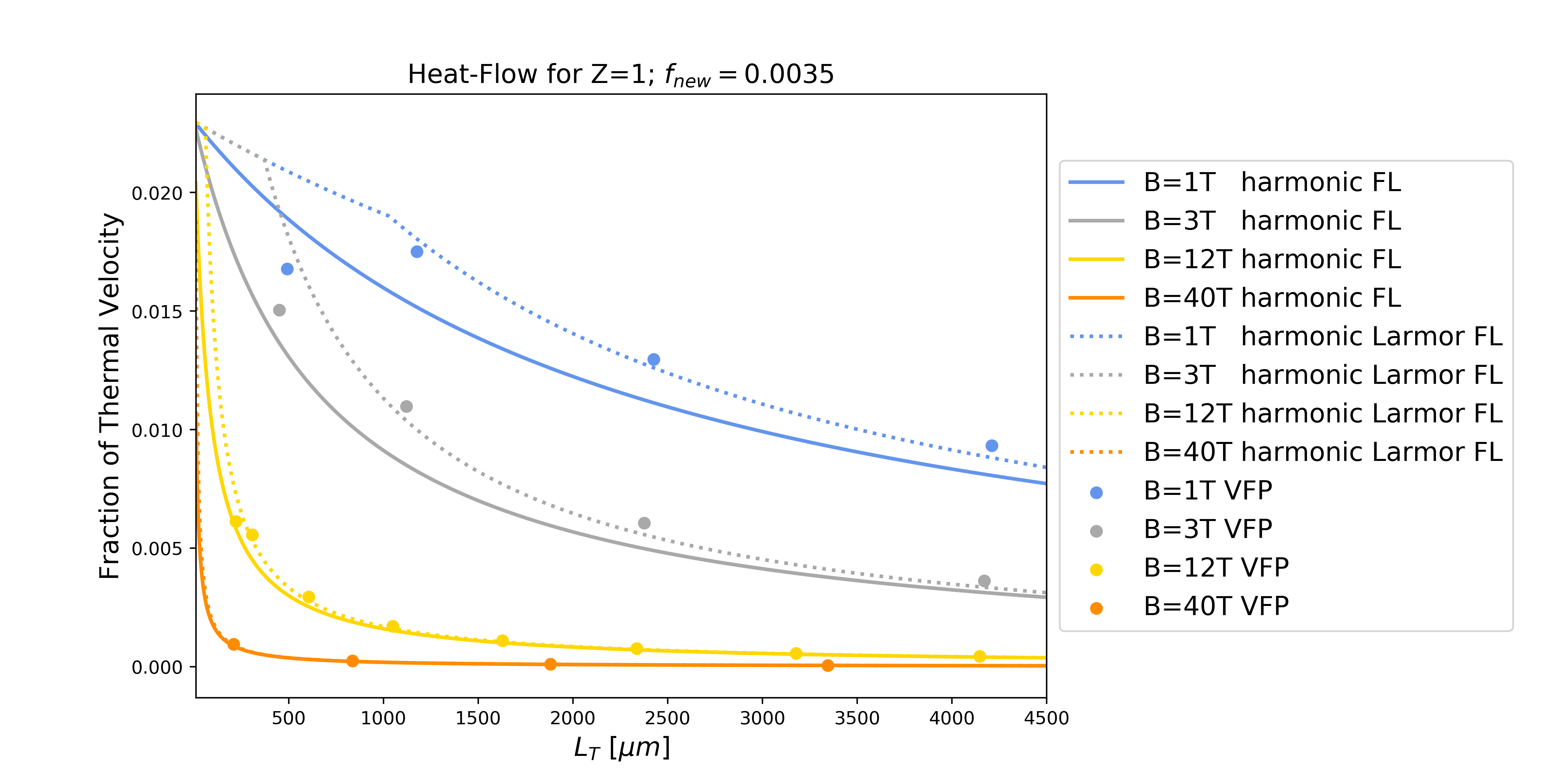}
		\caption{\label{fig:Harmonic_Larm_Z1}}
	\end{subfigure}
	
	\begin{subfigure}{\linewidth}
		\centering
		\includegraphics[scale=0.3]{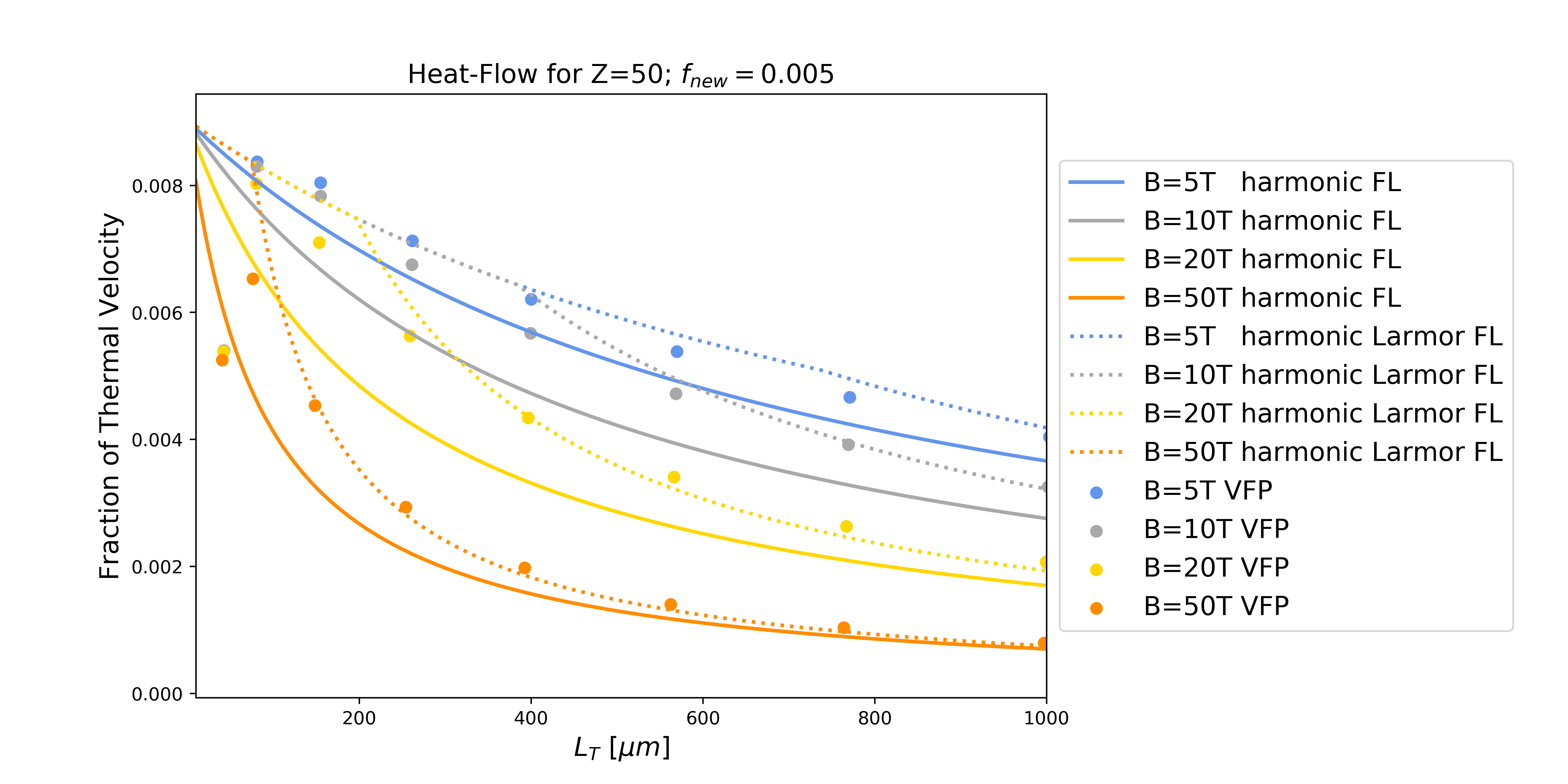}
		\caption{\label{fig:Harmonic_Larm_Z50}}
	\end{subfigure}
	\caption{Comparison of the standard harmonic flux-limiter (equation \ref{eq:q_harmonic_lambdad}) with a harmonic flux-limiter that has the collision length-scale adjusted to account for finite Larmor radius effects (equation \ref{eq:q_larm}). Top is for Z=1 and bottom for Z=50. Both plots show heat-flow normalized to the electron thermal velocity energy flux against temperature length-scale.  \label{fig:Harmonic_Larm} }	
\end{figure}

Figure \ref{fig:Harmonic_Larm} compares the harmonic flux limiter with the VFP data for various magnetic field strengths. The top figure is for Z=1 and the bottom is Z=50. 

The agreement between flux limiter theory and VFP data can be further improved by incorporating a magnetized length-scale $\lambda_B$ that accounts for the fact that the electron mean free path overestimates the non-locality when the Larmor radius is small:

\begin{equation}
\lambda_B = \frac{1}{\frac{1}{\lambda_d} + \frac{1}{r_L}} \label{eq:lamb_b} 
\end{equation}

Where $r_L = m_e v_{th}/(e|\underline{B}|)$ is the electron Larmor radius. 

The perpendicular heat-flow flux limiter then becomes:

\begin{equation}
|\underline{q}_{\bot Larmor}| = min \Bigg(\frac{\kappa_\bot |\nabla_\bot T_e |}{1 + \frac{\lambda_{B}}{f_{new} L_{T\bot}}\frac{\kappa_\bot^c}{\kappa_\parallel^c}} , |\underline{q}_{\bot Larmor}(|\underline{B}|=0)| \Bigg) \label{eq:q_larm}
\end{equation}

The improved agreement of equation \ref{eq:q_larm} can be seen in figure \ref{fig:Harmonic_Larm}. The magnetic field now has two competing effects, both from reducing the distance that the electron travels between collisions. First, the heat-flow is lower, as in the classical description, as their mobility is decreased. Second, the magnetic field is reducing the level of non-locality of the plasma, which increases the heat-flow. The VFP data always finds the magnetic field to overall reduce the heat-flow; the minimum function in equation \ref{eq:q_larm} is used to ensure this is replicated in the flux limiter. 

The impact of the Larmor radius adjustment is largest for moderate magnetic fields, where the Larmor radius is relatively small but the heat-flow has not been suppressed to a level that the electron transport is localized. 

It is tempting to see the agreement between harmonic flux limiters and the VFP data in figure \ref{fig:Harmonic_Larm} and assume that $f_{new}$ takes a value between 0.0035 and 0.0025 in all cases. Unfortunately the reality is not so simple. The VFP simulations were run using small amplitude single-mode perturbations; when the perturbations are non-linear, additional modes are introduced and no single flux limiter value can reproduce the results.


The final suggested modification to the flux limiter form is intended to replicate the decrease in heat-flow observed in VFP simulations at very low temperature length-scales. 

\begin{equation}
|\underline{q}_{\bot new}| = min \Bigg(\frac{\kappa_\bot |\nabla_\bot T_e |}{1 + \big(\frac{\lambda_{B}}{f_{new} L_{T\bot}}\frac{\kappa_\bot^c}{\kappa_\parallel^c} \big)^m} , |\underline{q}_{\bot new}(|\underline{B}|=0)| \Bigg) \label{eq:q_new}
\end{equation}

\begin{figure}
	\centering
	\begin{subfigure}{\linewidth}
		\centering
		\includegraphics[scale=0.3]{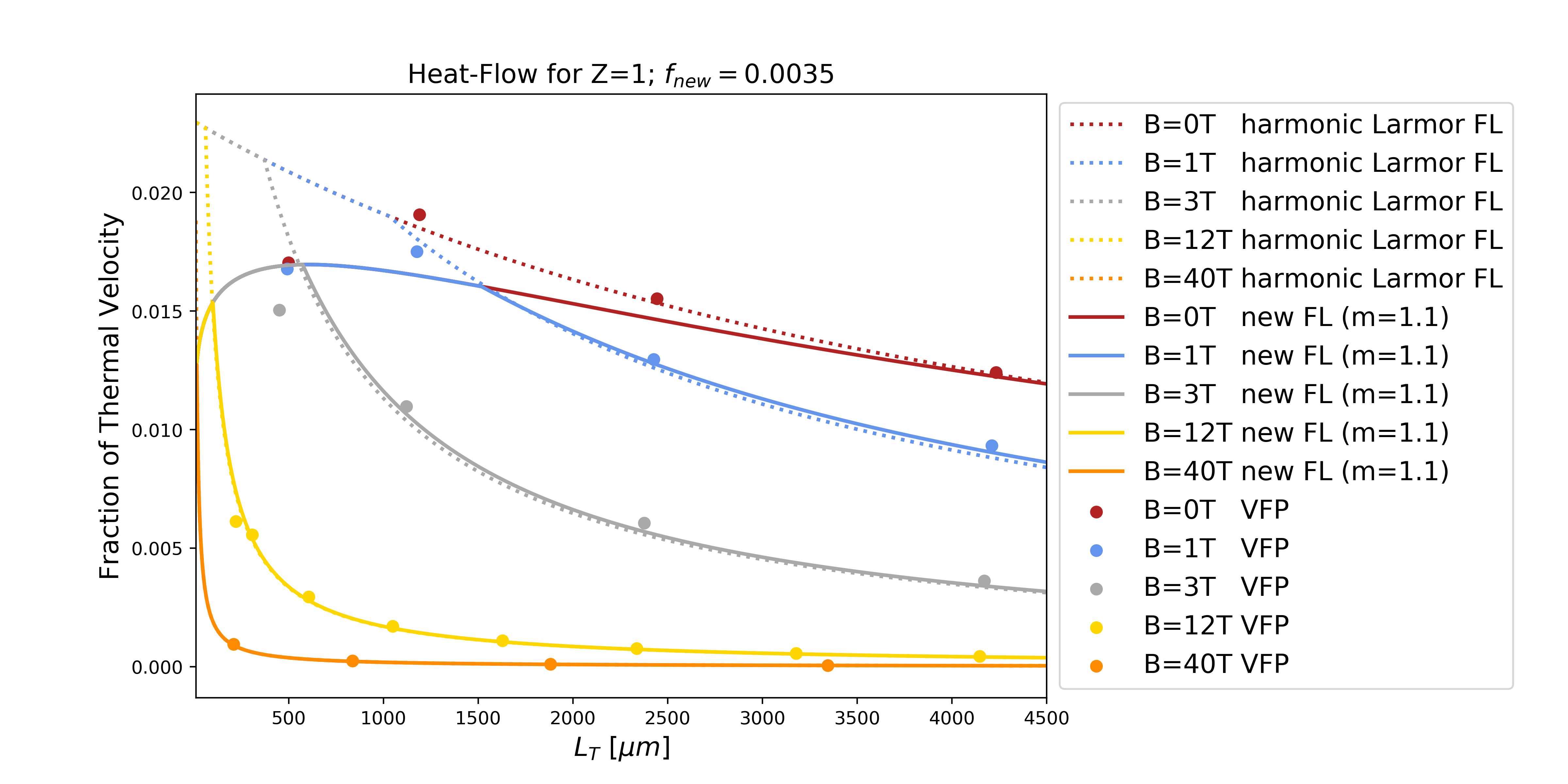}
		\caption{\label{fig:Harmonic_new_Z1}}
	\end{subfigure}
	
	\begin{subfigure}{\linewidth}
		\centering
		\includegraphics[scale=0.3]{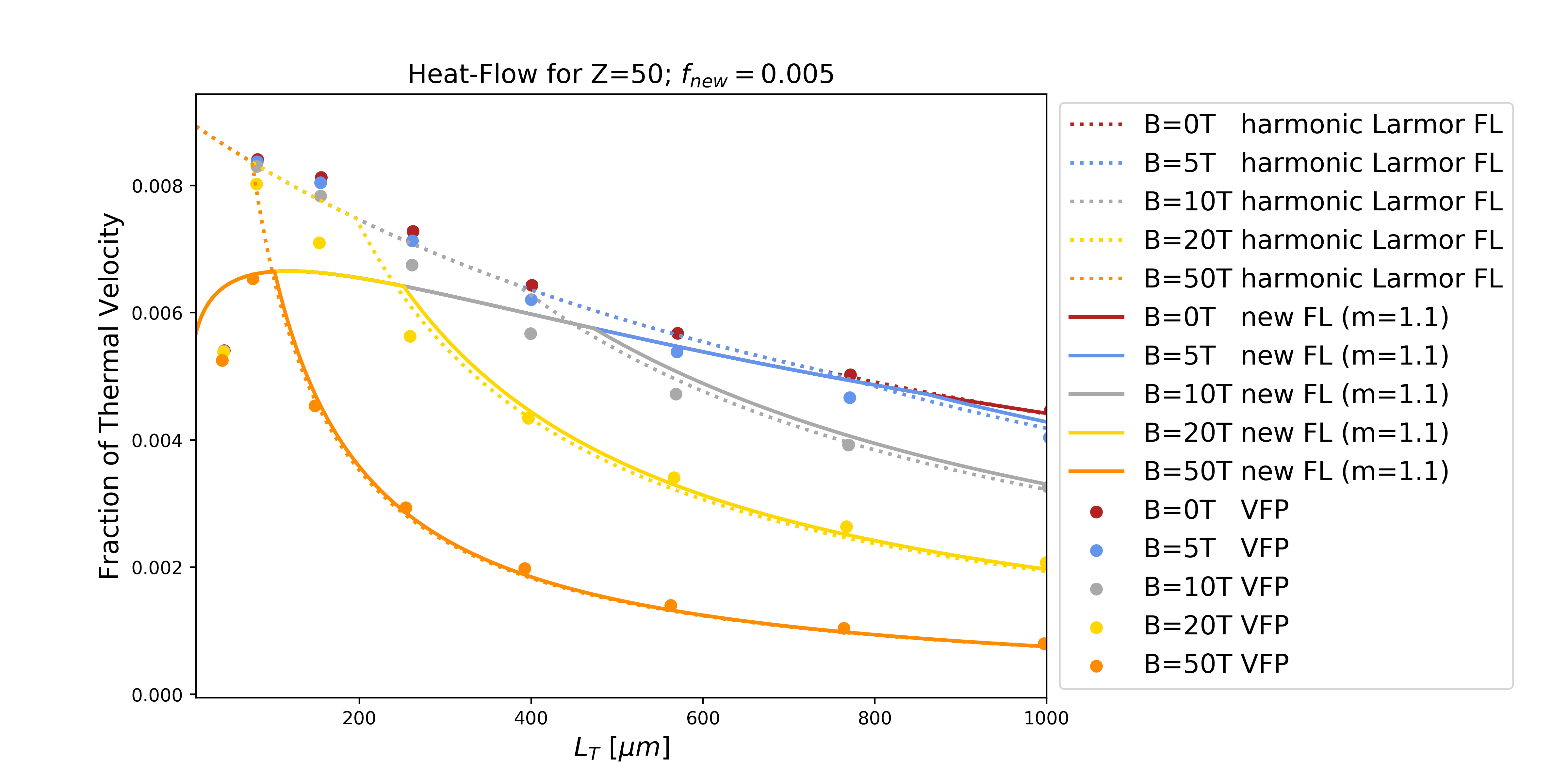}
		\caption{\label{fig:Harmonic_new_Z50}}
	\end{subfigure}
	\caption{ \label{fig:Harmonic_new} Comparison of the new flux limiter (equation \ref{eq:q_new}) using $m=1.1$ with the Larmor radius adjusted flux limiter (equation \ref{eq:q_larm}). Note that the latter flux limiter is the same as using the 'new' limiter with $m=1.0$. Using $m>1$ results in the heat-flow decreasing at low temperature length-scale, which is the behavior observed in VFP simulation.   }	
\end{figure}

Where $\lambda_{B}$ is the length-scale given by equation \ref{eq:lamb_b} and $m\ge 1$ is a parameter that is tuned to match either experimental or VFP data, much like $f_{new}$. Figure \ref{fig:Harmonic_new} compares this flux limiter to VFP data for $m=$1.1 and $m=$1.0 (i.e. the same as equation \ref{eq:q_larm}). The factor of 1.1 results in a decrease in the heat-flow at low temperature length-scale, but also deviates from the data at moderate length-scales. Of course, incorporating the additional free parameter $m$ into simulations has its drawbacks, particularly for predictive modeling of new experiments. However, it may also prove useful in explaining existing experimental conditions in the non-local regime.

\section{Nernst Flux Limiter \label{sec:nernst}}

The Nernst effect is an important mechanism in reducing the magnetization of a plasma by advecting magnetic fields down temperature gradients. Nernst and magnetized heat-flow are interconnected processes; heat-flow affects Nernst and Nernst affects the heat-flow, meaning that care must be taken to accurately represent the relative magnitude of each phenomenon in simulations. If the heat-flow is flux limited but not Nernst, the impact of magnetic fields will likely be underestimated. 

\begin{figure}
	\centering
	\begin{subfigure}{\linewidth}
		\centering
		\includegraphics[scale=0.3]{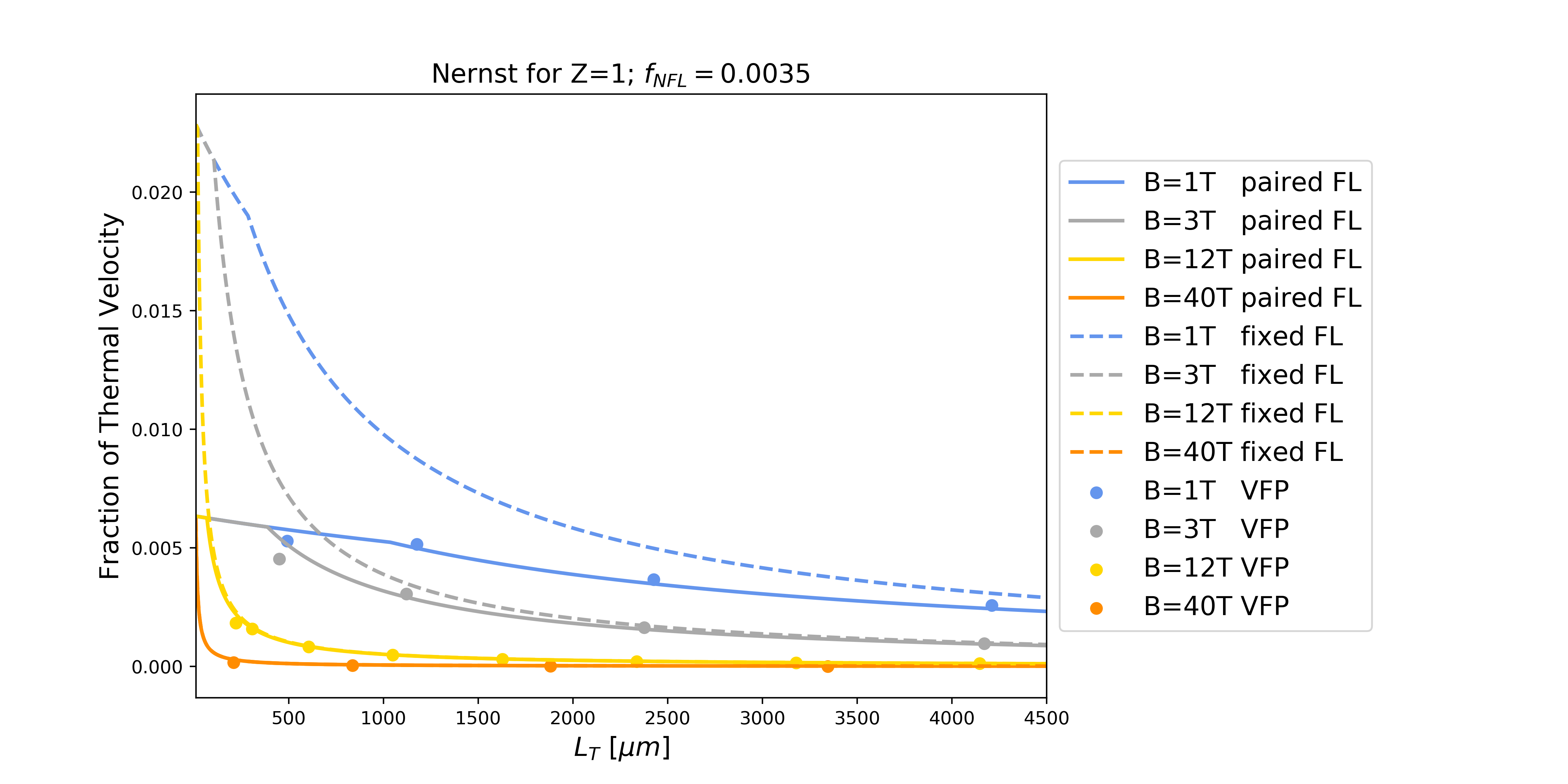}
		\caption{\label{fig:Harmonic_nernst_Z1}}
	\end{subfigure}
	
	\begin{subfigure}{\linewidth}
		\centering
		\includegraphics[scale=0.3]{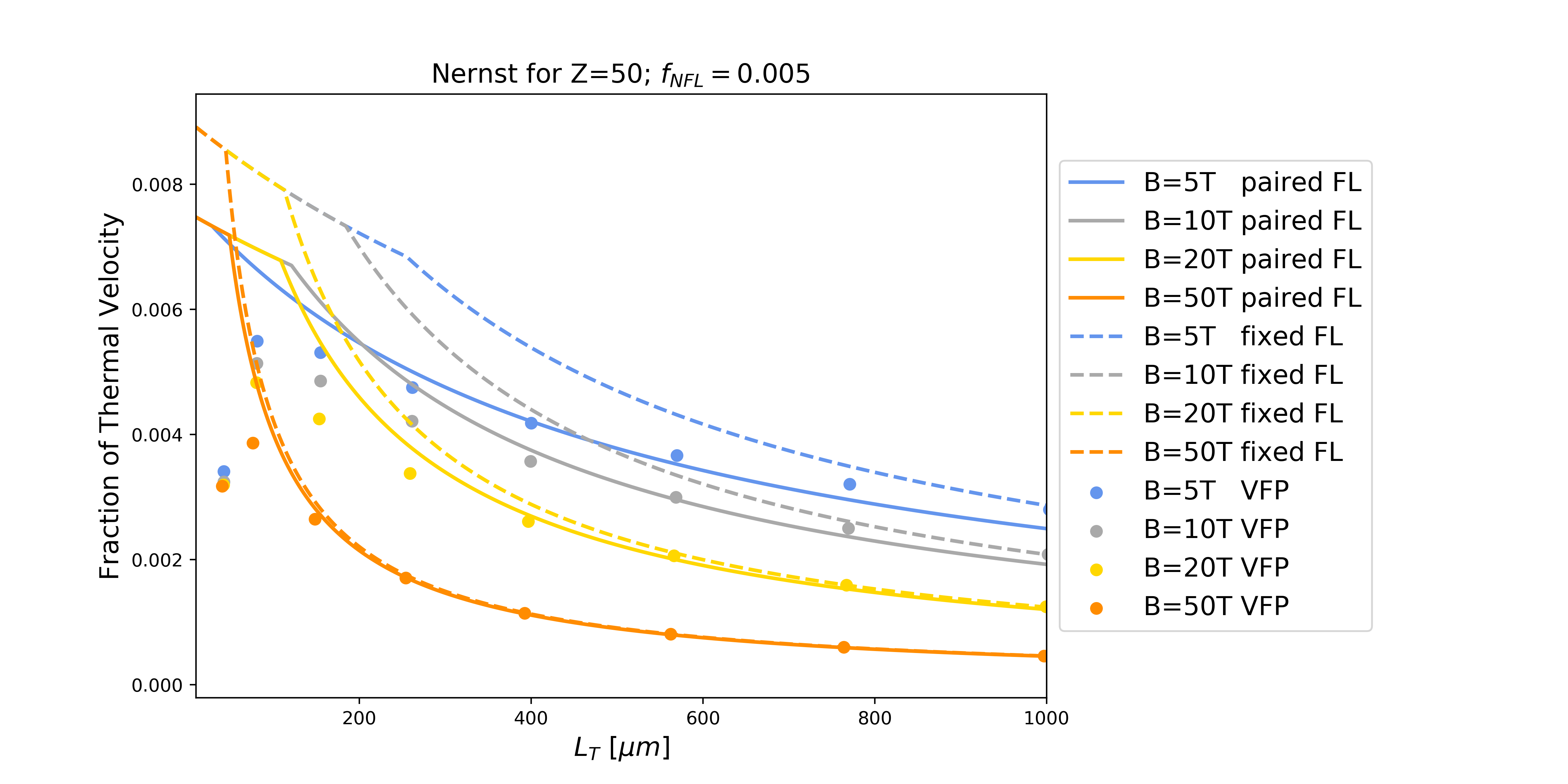}
		\caption{\label{fig:Harmonic_nernst_Z50} }
	\end{subfigure}
	\caption{ \label{fig:Harmonic_nernst} Nernst velocity as a fraction of electron thermal velocity for varying magnetic field strength and temperature length-scale. VFP simulations are denoted by dots, while flux limiter implementations are plotted as lines (with solid = paired limiter and dashed = fixed limiter). The top plot is for $Z=1$ and $f_{new} = 0.0035$. The bottom plot is for $Z=50$ and $f_{new} = 0.0025$. }	
\end{figure}

The transport of magnetic field in a classical plasma follows \cite{walsh2020}:
\begin{align}
\begin{split}
\frac{\partial \underline{B}}{\partial t} = & - \nabla \times \frac{\eta}{\mu_0 } \nabla \times \underline{B} + \nabla \times (\underline{v}_B \times \underline{B} ) \\
&+ \nabla \times \Bigg( \frac{\nabla P_e}{e n_e} - \frac{\beta_{\parallel} \nabla T_e}{e}\Bigg) \label{eq:mag_trans_new}
\end{split}
\end{align}
where the first term on the right-hand side is resistive diffusion with diffusivity $\eta$ and the final term is a source term. The second term represents the advection of magnetic fields with velocity $\underline{v_B}$:

\begin{equation}
\label{eq:mag_trans_new_velocity}	\underline{v}_B = \underline{v} - \gamma_{\bot} \nabla T_e - \gamma_{\wedge}(\underline{\hat{b}} \times \nabla T_e) 
\end{equation}
where $\underline{v}$ is the bulk plasma velocity, the second term is the Nernst term and the final term is the cross-gradient-Nernst term. The current-driven terms have been neglected from equation \ref{eq:mag_trans_new_velocity} due to their low impact in laser-driven plasmas \cite{10.1088/1361-6587/ac3f25}. The impact of electron kinetics on the source term in equation \ref{eq:mag_trans_new} will be the subject of a subsequent publication, although recent literature already suggests non-local corrections \cite{sherlock2020,davies2023}.

Note the likeness between the temperature gradient terms in equations \ref{eq:magheatflow} and \ref{eq:mag_trans_new_velocity}. The similarities between Nernst and heat-flow have long been a point of rumination \cite{haines1986,brodrick2018,walsh2020}. Brodrick \textit{et. al} were the first to notice that this similarity suggested the Nernst velocity should be flux limited in the same way the heat-flow has been limited in hydrocodes for almost 50 years. This section will propose possible models for implementation, demonstrate the importance of picking the correct model using Gorgon simulations, and will draw upon kinetic simulations to settle the debate.

The orthodox explanation of the heat-flow flux limiter is that it restricts the heat-flow velocity to be less than a certain fraction of the electron thermal velocity (as per equation \ref{eq:FL}). Using the same approach would suggest using:

\begin{equation}
	|\underline{v}_{N \bot}| = min\big(\gamma_{\bot} |\nabla T_e|, f_{FL} v_{th}\big) \label{eq:vN_naive}
\end{equation}

This approach ensures that the heat-flow velocity and the Nernst velocity have equivalent maximum magnitudes. This results in Nernst being limited fractionally less than the thermal heat-flow, as $\gamma_\bot^c < \kappa_\bot^c$ for all $Z$ and $\omega_e \tau_{ei}$. $\kappa^c/\gamma^c$ is approximately constant for varying $\omega_e \tau_{ei}$\cite{2021} and has a value of 3.63 for $Z=1$ and 1.20 for $Z=50$.

An alternate understanding comes from the recognition that thermal conduction of electron energy down temperature gradients and the advection of magnetic fields down temperature gradients (Nernst) are linked processes. As the ratio of classical transport coefficients $\kappa_\bot / \gamma_{\bot}$ is approximately constant for all plasma properties, it is possible that this underlying link is also maintained in the kinetic regime. 

Herein lies the central question of this section: should Nernst and heat-flow have the same maximum magnitude (when normalized by the electron thermal velocity) or the same fractional suppression? The former will be called a 'fixed' limiter, while the latter will be called a 'paired' limiter.

Section \ref{sec:heatflow} concluded that a harmonic flux limiter was necessary for matching VFP data; unsurprisingly, the same is found for Nernst. The Larmor radius correction to the electron collision distance is also found to improve accuracy of the Nernst limiter.

A 'paired' limiter, which ensures that the ratio of heat-flow to Nernst remains approximately constant (i.e. that they are being kinetically suppressed by the same fraction) is proposed:

\begin{equation}
|\underline{v}_{N \bot,paired}| = min \Bigg(\frac{\gamma_\bot \nabla T_e}{1 + \frac{\lambda_{B}}{f_{new} L_{T\bot}}\frac{\gamma_\bot^c}{\gamma_\parallel^c}} , |\underline{v}_{N \bot,paired}(|\underline{B}|=0)| \Bigg) \label{eq:vN_paired}
\end{equation}

Where $\gamma_\parallel^c = \gamma_\bot^c(\omega_e\tau_{ei}=0)$ and the length-scale $\lambda_{B}$ is given by equation \ref{eq:lamb_b}.

The alternate 'fixed' limiter takes the form:

\begin{equation}
|\underline{v}_{N\bot,fixed}| = min \Bigg(\frac{\gamma_\bot \nabla T_e}{1 + \frac{\lambda_{B}}{f_{new} L_{T\bot}}\frac{\gamma_\bot^c}{\kappa_\parallel^c}} , |\underline{v}_{N\bot,fixed}(|\underline{B}|=0)| \Bigg) \label{eq:vN_fixed}
\end{equation}
Where the only difference is the use of $\kappa_{\parallel}^c$ in the second equation.

To distinguish between the two Nernst flux limiter forms, the same VFP simulations from section \ref{sec:heatflow} for the thermal flux limiters can be used. Although the magnetic field is forced to be stationary in these simulations, the magnitude of Nernst is found by analyzing the electric field strength.

Figure \ref{fig:Harmonic_nernst} displays VFP data for the Nernst velocity as a function of temperature length-scale using various magnetic field strengths. Top shows $Z=1$ and bottom $Z=50$. The Nernst flux limiters have been set to the same value as for the heat-flow flux limiter ($f_{new} = 0.0035$ for $Z=1$ and $f_{new} = 0.0025$ for $Z=50$). Curves that show the different flux limiter models are plotted as lines (solid = paired limiter, dashed = fixed limiter). For both cases the paired limiter is closer to the VFP data, although this is much more obvious for $Z=1$.

Note that previous work by Brodrick \textit{et. al} agreed that the intrinsic connection between heat-flow and Nernst means that both should be reduced by the same fraction, i.e. the paired limiter is more appropriate.

Experiments to explicitly measure Nernst advection of magnetic fields down temperature gradients \cite{walsh2020} have been ongoing for a number of years now \cite{PhysRevLett.131.015101}, albeit not in the non-local regime. Ascertaining whether the paired limiter is correct form of Nernst flux limitation should be a priority in these experiments. 


\section{Cross Terms \label{sec:Xterms}}

\begin{figure}
	\centering
	\begin{subfigure}{\linewidth}
		\centering
		\includegraphics[scale=0.3]{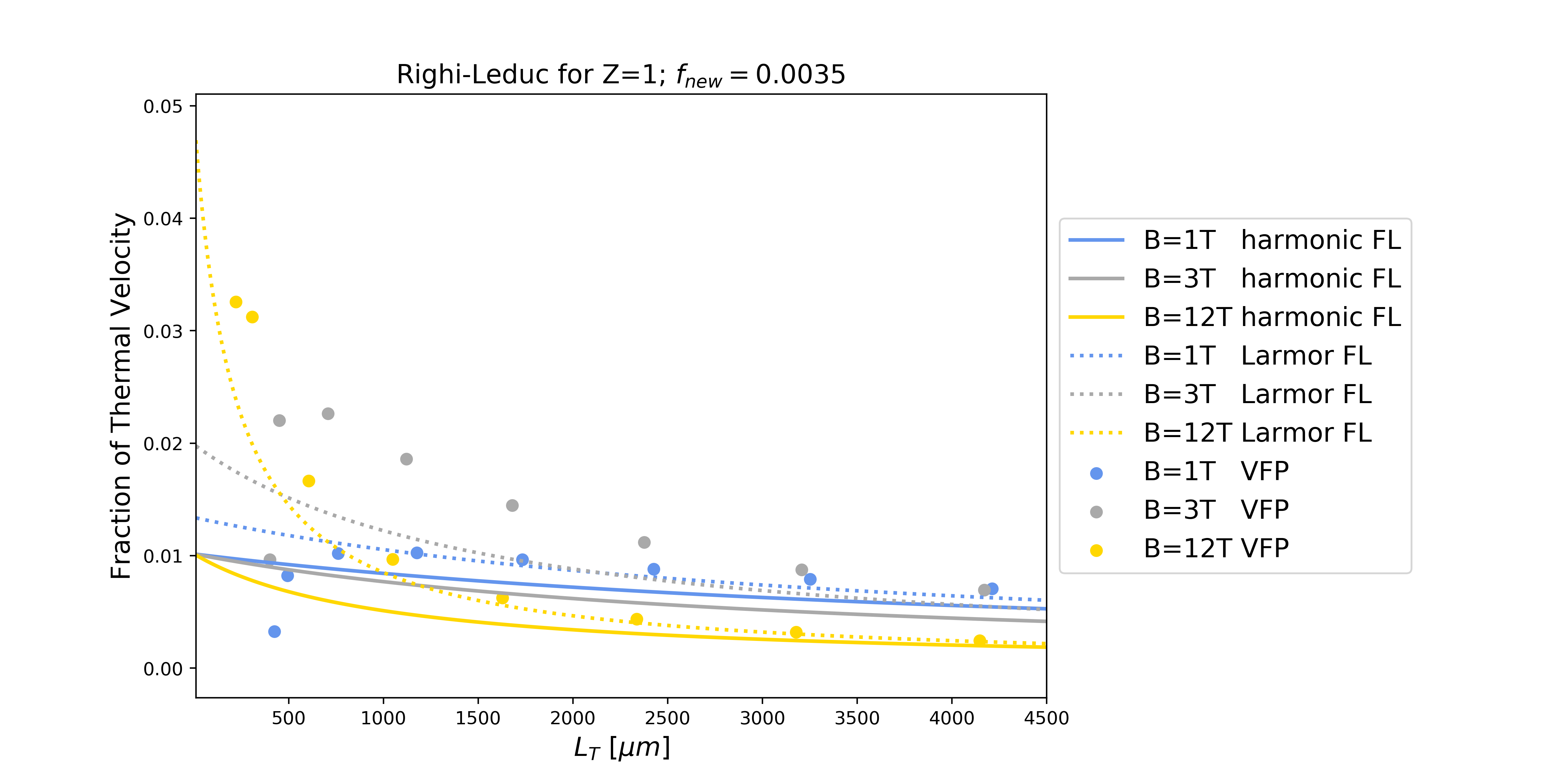}
		\caption{\label{fig:FLM_righi_Z1}}
	\end{subfigure}
	
	\begin{subfigure}{\linewidth}
		\centering
		\includegraphics[scale=0.3]{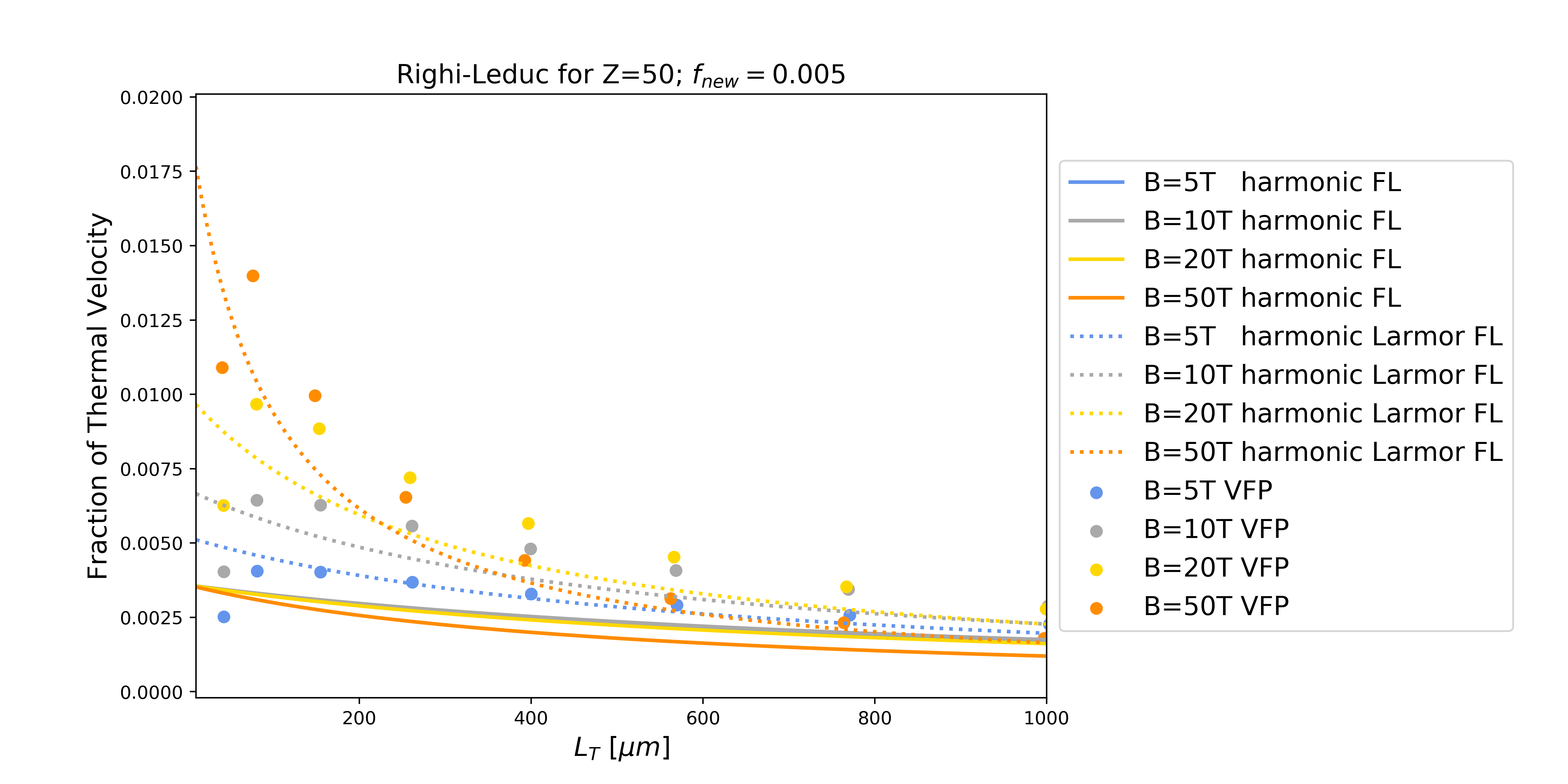}
		\caption{\label{fig:FLM_righi_Z50} }
	\end{subfigure}
	\caption{ \label{fig:FLM_righi} Righi-Leduc heat-flow as a fraction of thermal velocity energy flux, showing that the Larmor radius flux-limiter correction brings agreement closer to VFP data. Top is for Z=1, bottom for Z=50.}	
\end{figure}

This section deals with the final terms in equations \ref{eq:magheatflow} and \ref{eq:mag_trans_new_velocity}. These represent the deflection of heat-flow and magnetic field due to partial Larmor orbits between collisions. The former is called the Righi-Leduc heat-flow, while the latter is called the cross-gradient-Nernst term. As in the case of the perpendicular advection of magnetic field and electron energy, these terms are intrinsically  linked.

The coefficient $\gamma_{\wedge}$ should be calculated from \cite{sadler2021} or \cite{davies2021} as the coefficients given by Epperlein \& Haines were found to be erroneous, giving too large cross-gradient-Nernst velocities at low magnetization \cite{2021}.

It will be shown here that the Larmor radius correction is essential to calculating the Righi-Leduc heat-flow in the non-local regime. 

The following flux limiter for Righi-Leduc heat-flow is proposed:

\begin{equation}
|\underline{q}_{\wedge Larmor}| = \frac{-\kappa_\wedge \underline{\hat{b}}\times\nabla T_e}{1 + \Big( \frac{\lambda_{B}}{f_{new} L_{T\bot}}\frac{\kappa_\wedge^c}{\kappa_{\wedge,max}^c}\Big)^m}  \label{eq:righi_larm}
\end{equation}

Where $\kappa_{\wedge,max}^c$ is the maximum value of non-dimensional Righi-Leduc coefficient for a given plasma ionization. A function for finding this value is useful for code implementation and is given in section \ref{sec:Code} under equation \ref{eq:kappa_max}.

Figure \ref{fig:FLM_righi} compares the flux limiter form of equation \ref{eq:righi_larm} using $m=1.0$ against VFP data for the same conditions as discussed in the prior sections. A case that does not use the Larmor length-scale correction is also included, deviating more from the kinetic data. 

\begin{figure}
	\centering
	\centering
	\includegraphics[scale=0.5]{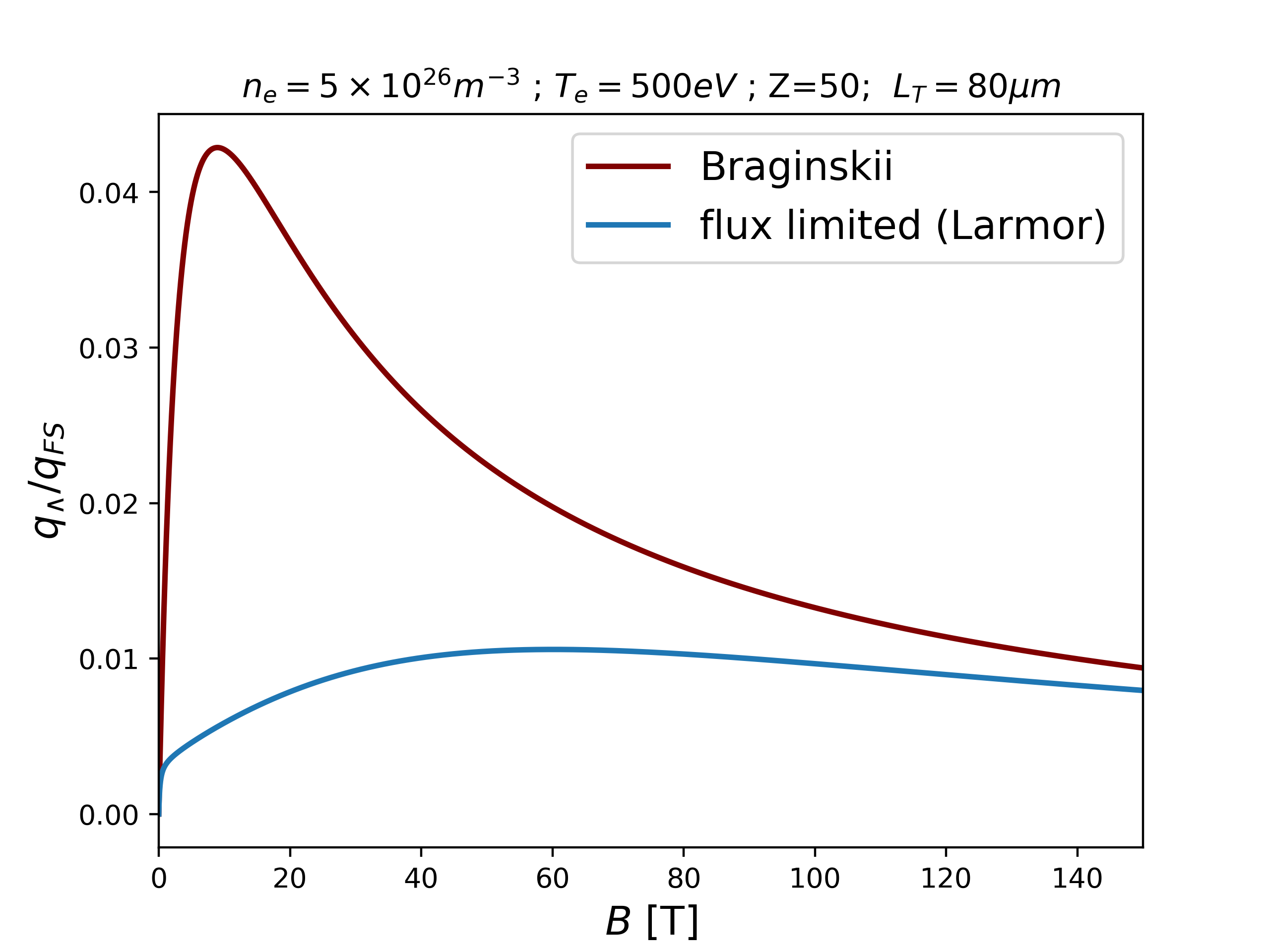}\caption{ \label{fig:Righi_theory} Theoretical Righi-Leduc heat-flow as a fraction of free-streaming against magnetic field strength, without a flux-limiter (Braginskii) and with the Righi-Leduc flux limitation suggested by equation \ref{eq:righi_larm}. The peak heat-flow moves to higher magnetic field strength when flux-limited. A magnetic field of 10T corresponds to a Hall Parameter of $\omega_e \tau_e =0.13$ in these conditions.}	
\end{figure}

One side-effect of flux limiting the Righi-Leduc heat-flow is that the peak heat-flow occurs at a larger magnetic field strength. Figure \ref{fig:Righi_theory} shows the classical Righi-Leduc heat-flow (Braginskii) against magnetic field strength, with the characteristic peak at low magnetic field strength (9T). The plasma conditions are given at the top of the figure. Once the flux-limiter is included, the magnetic field strength helps to reduce the level of non-locality, resulting in a peak Righi-Leduc heat-flow at 60T.

\begin{figure}
	\centering
	\begin{subfigure}{\linewidth}
		\centering
		\includegraphics[scale=0.3]{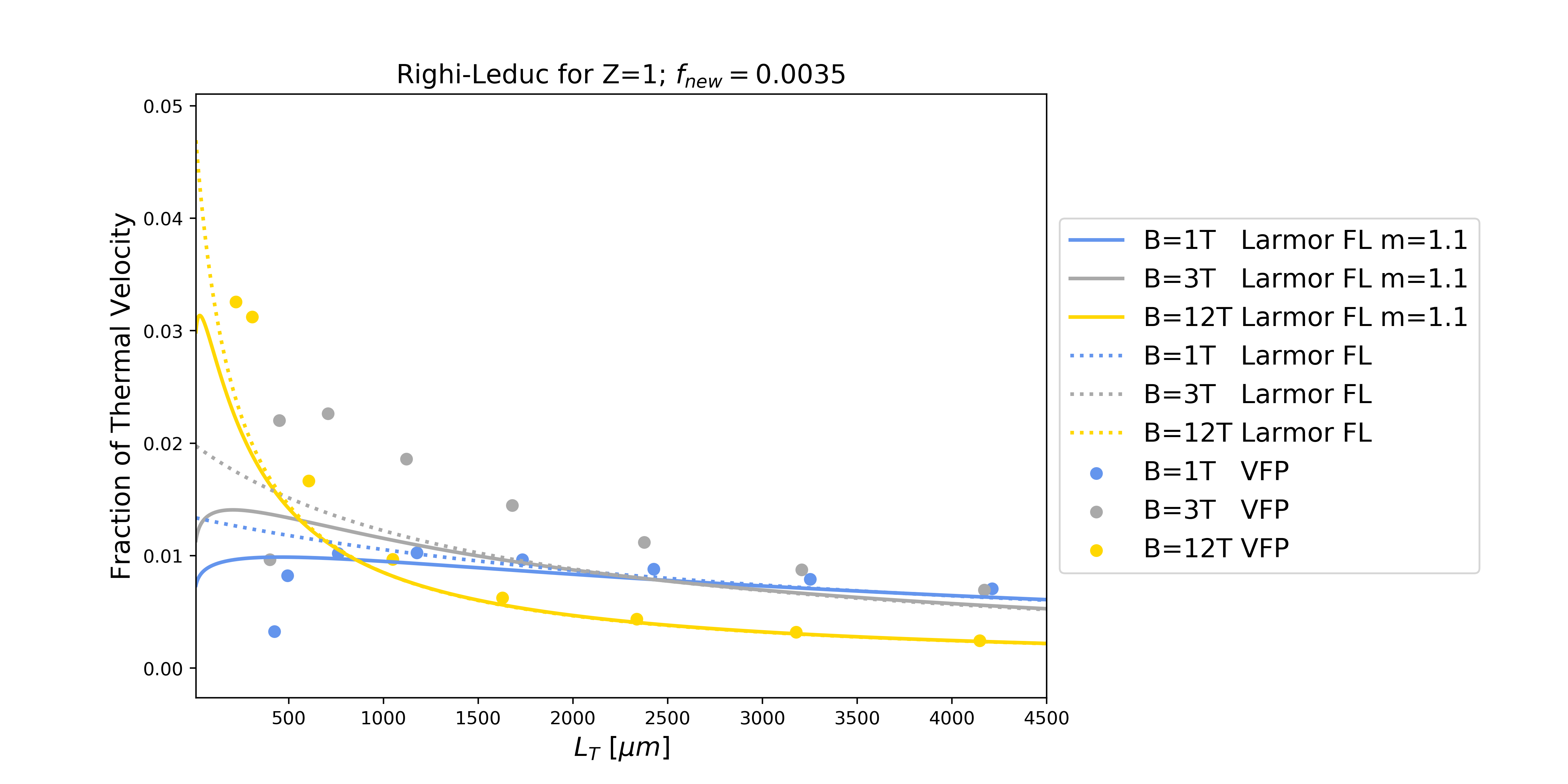}
		\caption{\label{fig:FLM_righi_new_Z1} }
	\end{subfigure}
	
	\begin{subfigure}{\linewidth}
		\centering
		\includegraphics[scale=0.3]{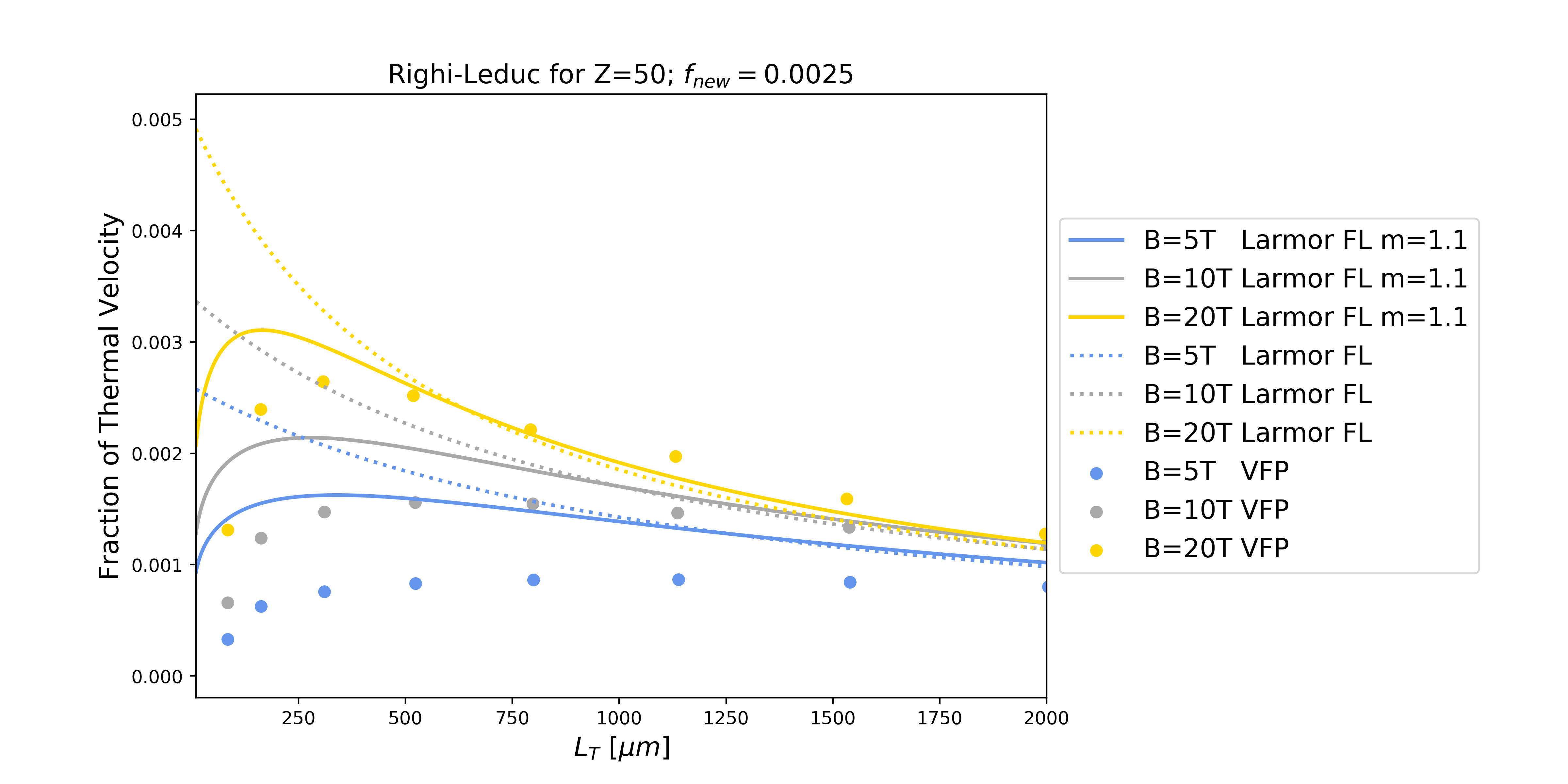}
		\caption{\label{fig:FLM_righi_new_Z50} }
	\end{subfigure}
	\caption{ \label{fig:FLM_righi_new} Righi-Leduc heat-flow as a fraction of free-streaming thermal flux to show the impact of using $m>1$.}	
\end{figure}

The agreement with VFP data can be somewhat improved by using $m>1$ in equation \ref{eq:righi_larm}. Figure \ref{fig:FLM_righi_new} demonstrates this for $m=1.1$, which is the value that gave the best agreement for perpendicular heat-flow in figure \ref{fig:Harmonic_new}. For Righi-Leduc, however, a value of $m=1.2$ is found to give better agreement. This is likely due to the fact that the cross-terms depend on a higher moment of the electron distribution, which results in a higher level of kinetic suppression. 

Similarly, the following flux limitation of the cross-gradient-Nernst term is proposed:
\begin{equation}
|\underline{v}_{N\wedge Larmor}| = \frac{-\gamma_\wedge \underline{\hat{b}}\times\nabla T_e}{1 + \Big( \frac{\lambda_{B}}{f_{new} L_{T\bot}}\frac{\gamma_\wedge^c}{\gamma_{\wedge,max}^c}\Big)^m}  \label{eq:cross}
\end{equation}

\begin{figure}
	\centering
	\begin{subfigure}{\linewidth}
		\centering
		\includegraphics[scale=0.3]{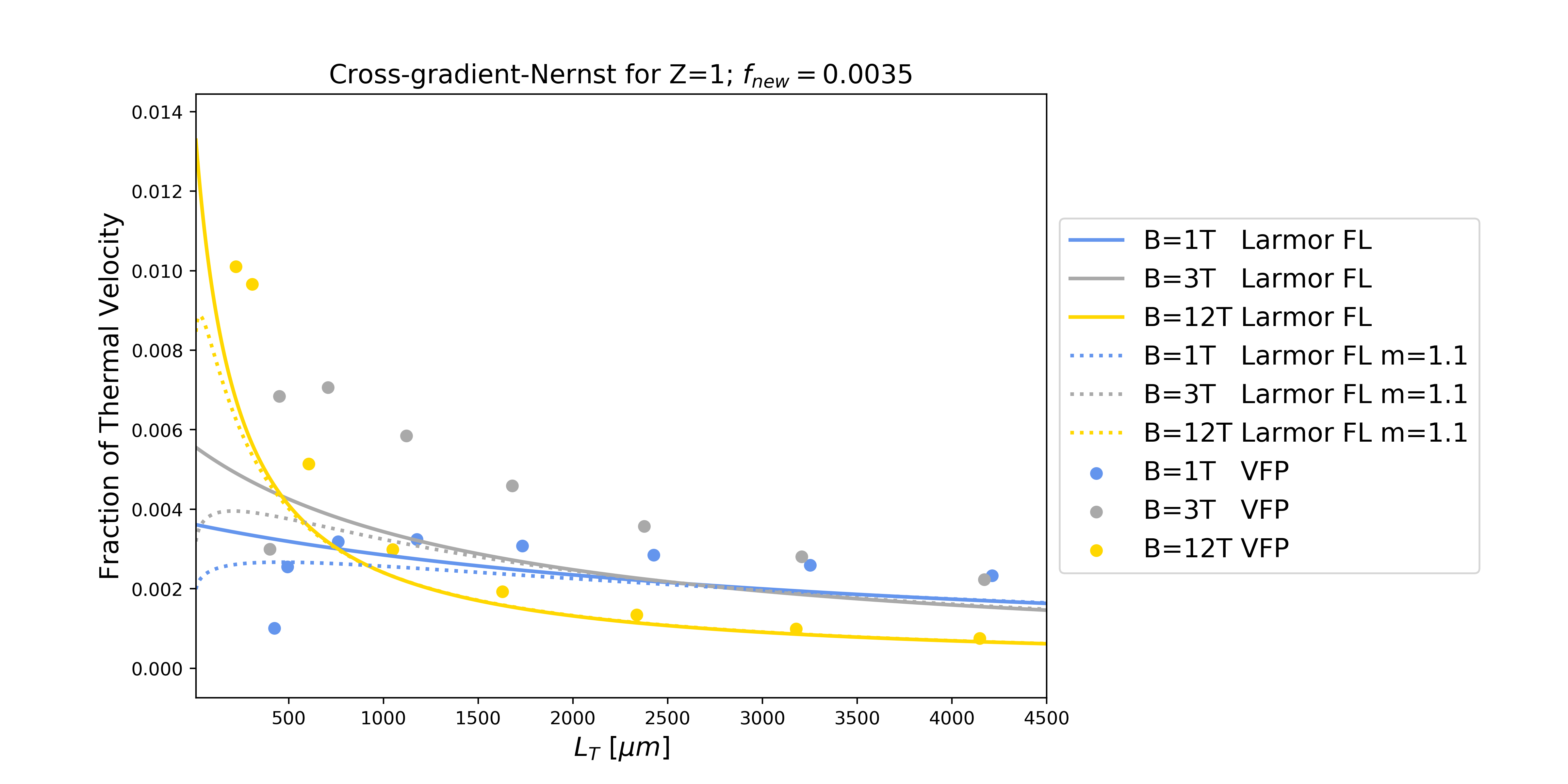}
		\caption{\label{fig:FLM_xnernst_Z1}}
	\end{subfigure}
	
	\begin{subfigure}{\linewidth}
		\centering
		\includegraphics[scale=0.3]{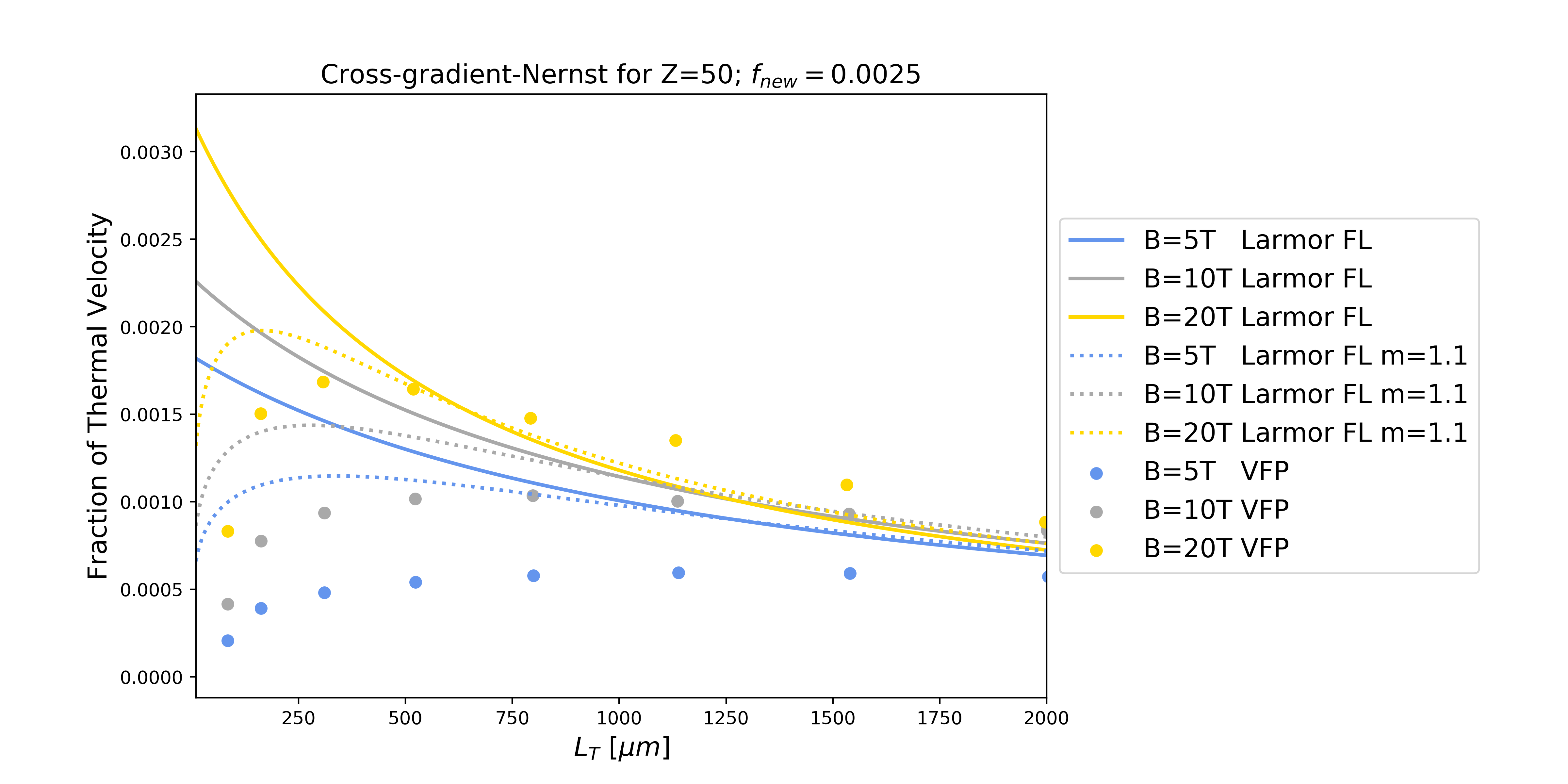}
		\caption{\label{fig:FLM_xnernst_Z50} }
	\end{subfigure}
	\caption{ \label{fig:FLM_xnernst} Cross-gradient-Nernst velocity as a fraction of the electron thermal velocity. The cross-gradient-Nernst term requires similar flux limitation to the Righi-Leduc heat-flow in order to reproduce kinetic results. }	
\end{figure}

Figure \ref{fig:FLM_xnernst} compares with VFP data for $m=1.0$ and $m=1.1$, with the data clearly showing the link between Righi-Leduc heat-flow and cross-gradient-Nernst is maintained in the kinetic regime.

\section{Non-local corrections to unmagnetized heat-flow in a hohlraum \label{sec:NoMHD_impact}}

\begin{figure}
	\centering
	\begin{subfigure}{\linewidth}
		\centering
		\includegraphics[scale=0.7]{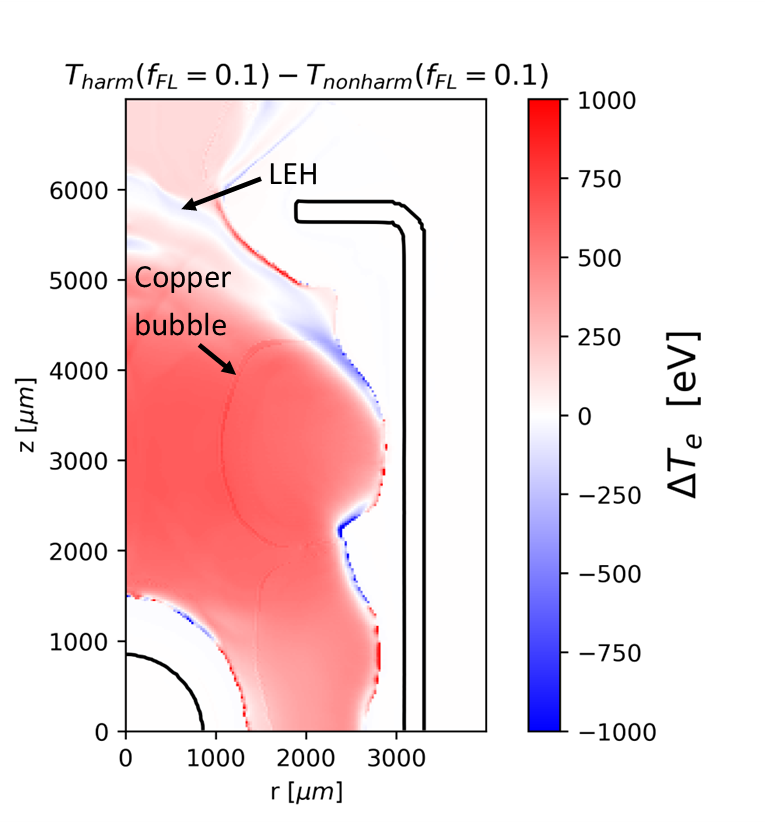}
		\caption{\label{fig:hohlraum_nomhd1}}
	\end{subfigure}
	
	\begin{subfigure}{\linewidth}
		\centering
		\includegraphics[scale=0.7]{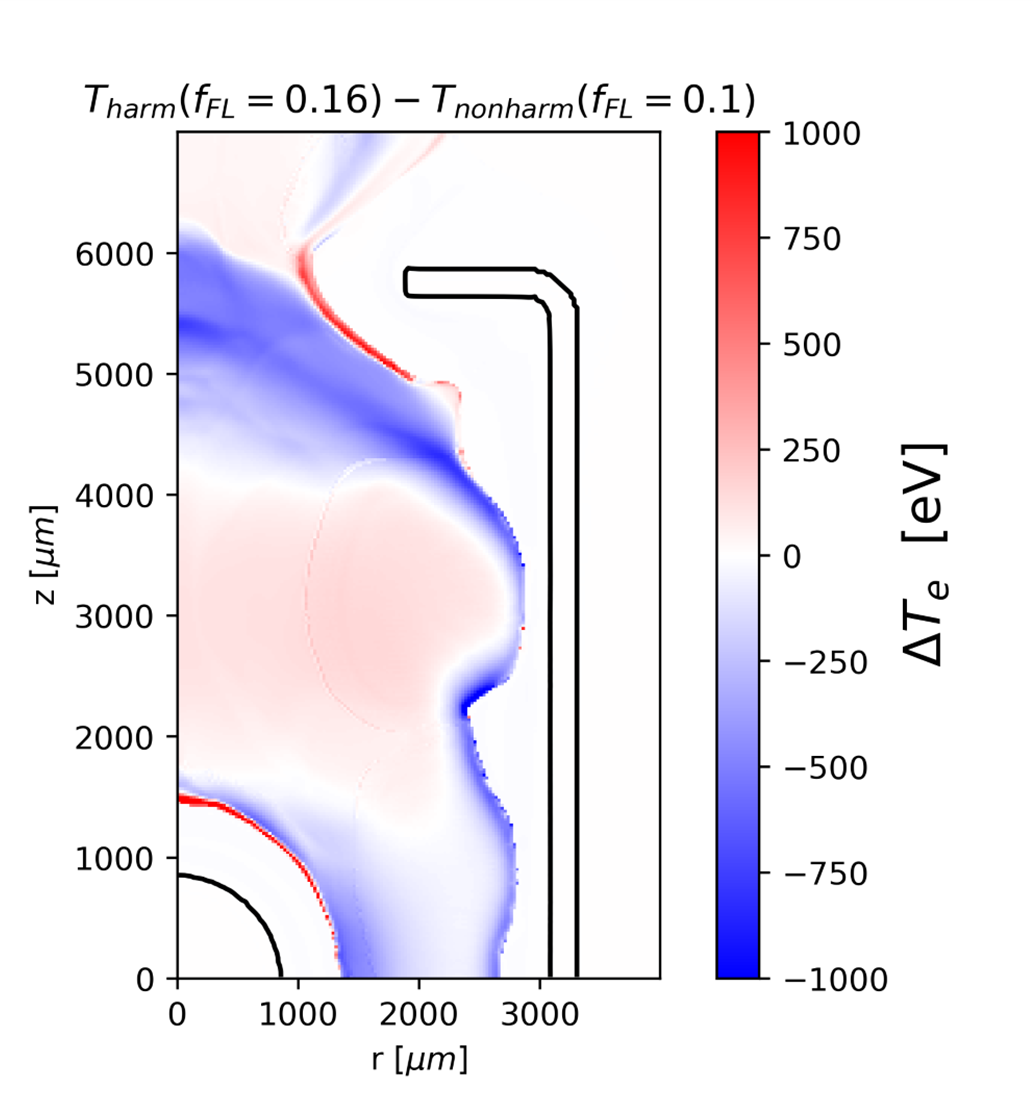}
		\caption{\label{fig:hohlraum_nomhd2}}
	\end{subfigure}
	\caption{ \label{fig:hohlraum_nomhd} Change in simulated hohlraum electron temperature by moving from a non-harmonic flux limiter to a harmonic flux limiter. No MHD was used in any of the cases. Top uses a harmonic flux limiter value of $f_{FL}=0.1$ and bottom uses a value of $f_{FL}=0.16$. In both cases the non-harmonic flux limiter value is $f_{FL}=0.1$. Black lines denote regions of solid density material, showing the hohlraum and capsule outline. The value of $f_{FL}=0.16$ in the lower case was chosen to roughly replicate the copper bubble temperature of the non-harmonic case.}	
\end{figure}

As the vast majority of design calculations for ICF experiments are conducted without accounting for naturally occurring magnetic fields, it is pertinent to first demonstrate the thermal flux limiter modifications without MHD included (i.e. only using the flux limiters in section \ref{sec:heatflow_noB}). The two main changes from the orthodoxy are using a harmonic mean (rather than hard cut-off) and using the non-locality length-scale (rather than electron ion mean free path).

To demonstrate the impact of these changes, a set-up relevant to a NIF hohlraum is used. NIF hohlraums are notoriously kinetic in nature and much experimental effort has been put into finding the correct flux limiter for rad-hydro calculations \cite{glenzer1997,barrios2018,farmer2018}.  The use of 2 different plasma species (hohlraum material and gas fill) is also convenient for elucidating the impact of the non-locality parameter giving a different scaling of flux limit with plasma ionization.

The Gorgon simulations use a copper hohlraum filled with 0.3mg/cc 4He. Copper is used here due to lack of access to a quality gold opacity model, but provides a reasonable surrogate for elucidating key Physics. The inner radius of the hohlraum is 3.1mm with a laser entrance hole (LEH) of radius 1.9mm. A capsule of radius 844$\mu$m is deployed at the center of the hohlraum, which ablates throughout the simulation from the radiation drive. The capsule is made of solid copper in this case, due to limitations in the code for incorporating more than 2 plasma species. While this is not a realistic ICF target, the non-local Physics of interest is less important around the capsule. No hohlraum window is used in these simulations.

The 1.4MJ laser drive is simulated using 3-D ray tracing and inverse Bremmstrahlung absorption. The beams are split into two groups: 'inners' and 'outers'. The inner beams are incident on the hohlraum wall close to the capsule waist, while the outers strike closer to the LEH. The temperature and expansion velocity of the gold/copper bubble caused by the outer beams is of particular interest, as it has the greatest impact on capsule drive energetics and symmetry \cite{ralph2018}. All simulation outputs are analyzed at t=6ns, which is during the peak laser intensity.

The 2-D Gorgon simulations use a cylindrical mesh with reflective boundary conditions at z=0, i.e. at the capsule waist. The mesh resolution is 10$\mu$m radially and axially. The radiation transport used is non-diffusive, using a P1/3 flux limiter \cite{jennings2006} (not related to the thermal flux limiter). Multi-group non-LTE opacities and emissivities are used. 

 \begin{figure}
 	\centering
 	\centering
 	\includegraphics[scale=0.7]{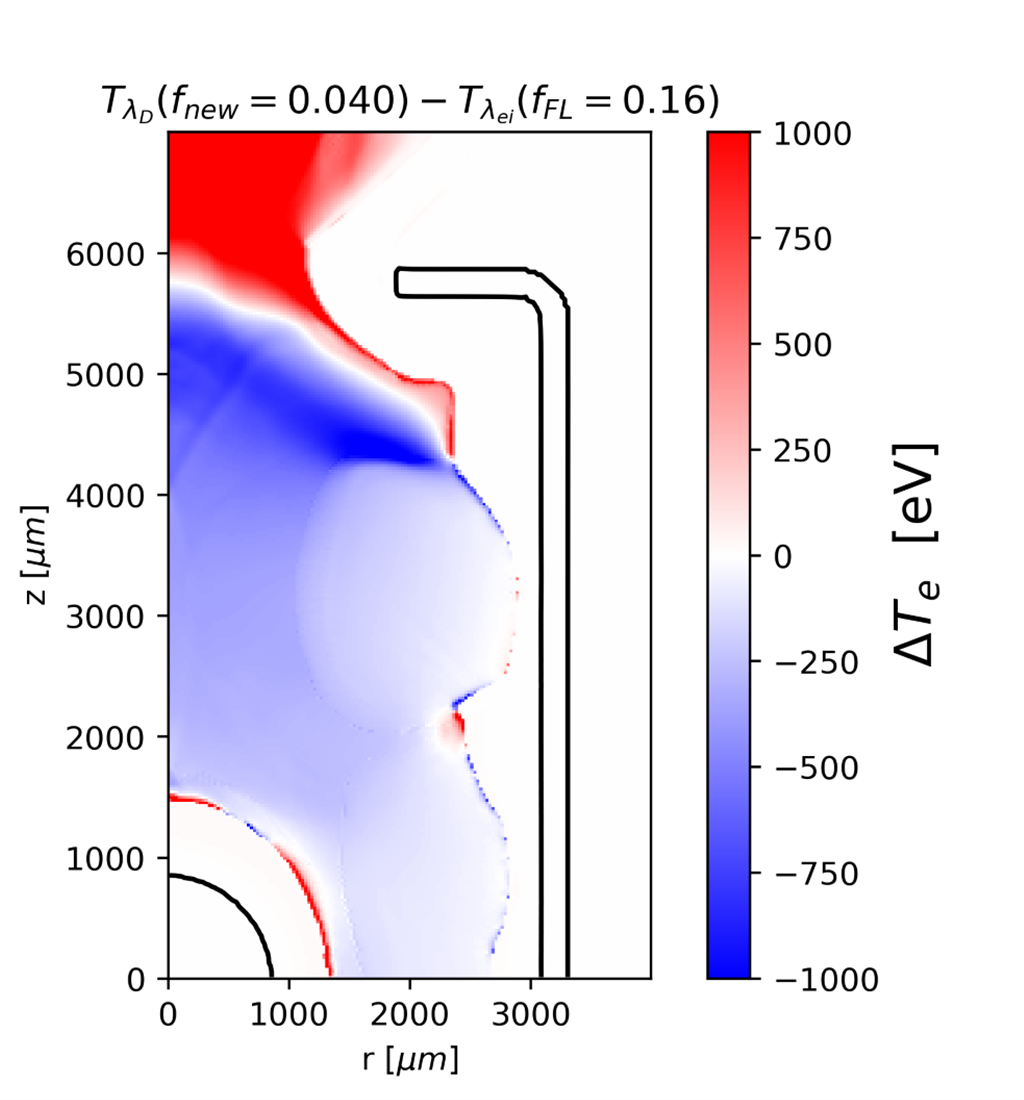}\caption{ \label{fig:hohlraum_lambda} This figure demonstrates the importance of how the chosen flux limiter varies with plasma ionization. The color is change in electron temperature in a  hohlraum when moving from a flux limiter using the mean free path to the non-locality length-scale. No magnetic fields were included in these calculations and both cases use a harmonic style limiter. The value of $f_{new}=0.040$ was chosen to give approximately the same copper bubble temperature. }	
 \end{figure}
 
Figure \ref{fig:hohlraum_nomhd1} shows the change in electron temperature in the hohlraum moving from a non-harmonic flux limiter (equation \ref{eq:q_nonharmonic}) to a harmonic version (equation \ref{eq:q_harmonic}), while keeping the flux limiter constant at $f_{FL} = 0.1$. The flux limiter value has been chosen to be in the range typically used by ICF design codes. The black lines indicate solid density material, showing the outline of the hohlraum wall and the capsule. Annotations of the LEH (where the laser enters into the hohlraum) and the edge of the copper bubble pushing into the underdense helium are shown. Clearly the harmonic flux limiter is more restrictive, increasing the temperature by around 600eV in most of the hohlraum plasma. This is not surprising, as figure \ref{fig:FLM_unmag} shows the harmonic heat-flow always being lower than the non-harmonic alternative. 

The value of the flux limiter is somewhat arbitrary, being tuned to approximate experiment data. The harmonic flux limiter value does not have to be the same as the non-harmonic. In figure \ref{fig:hohlraum_nomhd2} a harmonic flux limiter value of $f_{FL}=0.16$ is used, which was found to get the closest temperature to the non-harmonic case in the copper bubble. Nonetheless, differences are observed, particularly lowering the temperature at the higher density hohlraum wall and in the LEH helium plasma.

Next, the impact of using the non-locality length-scale in the thermal flux limiter is studied (i.e. comparing the use of equation \ref{eq:q_harmonic} with \ref{eq:q_harmonic_lambdad}). A value of the updated flux limiter of $f_{new} = 0.04$ was found to give the closest temperature in the copper bubble. Figure \ref{fig:hohlraum_nomhd2} shows the change in temperature. While the bubble is the same temperature, the helium fill is somewhat cooler, although the plasma outside the LEH is substantially hotter (around 1keV).

Figure \ref{fig:FLM_vs_Z} showed the dependence of the mean free path limiter and the non-locality length-scale limiter against Z. The value $f_{new}=0.04$ gave the same flux limitation for $Z\approx 25$. With copper having a maximum ionization state of 29, it is clear that both the harmonic flux limiter with $f_{FL}=0.1$ and the non-locality length-scale limiter with $f_{new}=0.04$ give similar heat-flow in the copper bubble. 

While a flux limiter can be found to give the same result in a chosen plasma species, the restriction on heat-flow will be different elsewhere. The thermal conduction in the helium using the new flux limiter form is more restricted than the old form (again see figure \ref{fig:FLM_vs_Z}). This results in the $> 1keV$ change in electron temperature outside the LEH. Inside the LEH, however, the helium temperature is largely set by the background radiation field, which is set by the copper wall. 

A value of $m>1$ was simulated, which more closely reproduces the VFP data at very low thermal length-scales. However, the impact was minimal, with less than 200eV changes in temperature observed.

\section{Non-local corrections to magnetized heat-flow and Nernst in a hohlraum \label{sec:MHD_impact}}

\begin{figure}
	\centering
	\centering
	\includegraphics[scale=0.7]{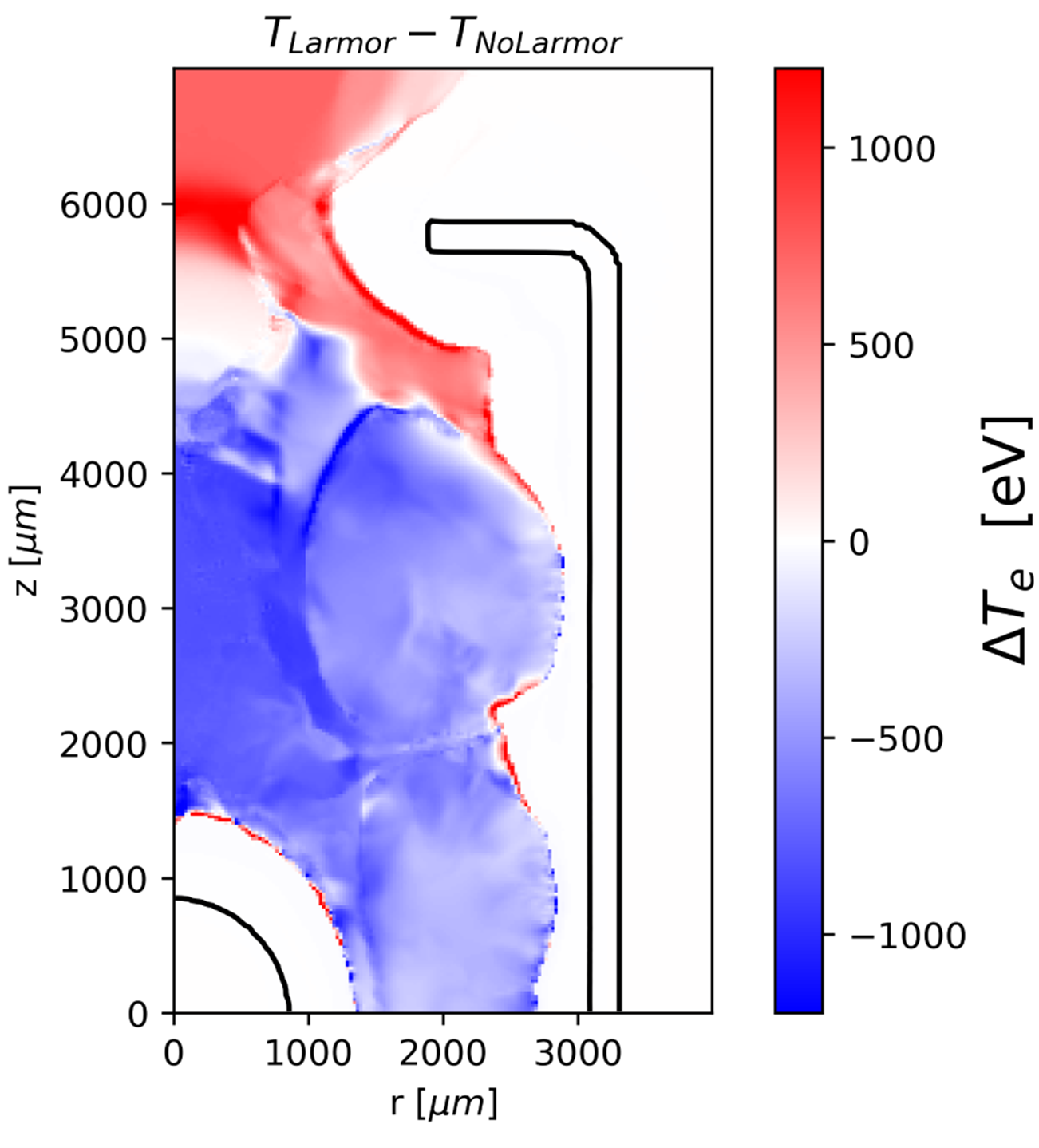}\caption{ \label{fig:hohlraum_Larmor} Change in electron temperature in a hohlraum simulation with MHD when a correction to the electron collision length-scale due to finite Larmor orbits is included in the flux limiter (see equation \ref{eq:lamb_b}). While the self-generated magnetic fields reduce the electron heat-flow overall, this figure shows the contribution to increasing the heat-flow by reducing the level of non-local flux limitation.}	
\end{figure}

This section demonstrates the importance of non-local corrections to magnetized heat-flow and Nernst in hohlraum simulations that include self-generated magnetic fields. The set-up is the same as discussed in section \ref{sec:NoMHD_impact}, but with the magnetic flux source terms included (the final term in equation \ref{eq:mag_trans_new}). Biermann Battery is the dominant source term, although there is also a component generated by gradients in plasma ionization \cite{sadler2020}. All cases in this section use a flux limiter value of $f_{new} = 0.04$

This paper has focused on non-local corrections to the heat-flow and Nernst. It is also known that the source of magnetic flux is suppressed in kinetic conditions \cite{sherlock2020}. To date, 2 models for that suppression have been suggested: the first by Sherlock \& Bissell \cite{sherlock2020}; the second by Davies \cite{davies2023}. The work in this section uses the model proposed by Sherlock \& Bissell, as it has been somewhat validated against proton radiographs of laser-driven foil targets \cite{campbell2021measuring}. A thorough comparison of the different proposed models will be the subject of a subsequent publication.

	
 
Self-generated magnetic fields in hohlraums have been studied previously using extended-MHD codes \cite{farmer2017}, finding a subtle balance between Biermann Battery magnetizing the plasma and the Nernst effect demagnetizing the plasma. The intricacies of this system will not be discussed in great detail here, as the focus is on showing the importance of the different flux limiter implementations suggested in this paper. 

Figure \ref{fig:hohlraum_Larmor} shows the change in simulated electron temperature through including the Larmor radius correction to the mean free path from equation \ref{eq:lamb_b}. The self-generated magnetic field is resulting in the plasma being less kinetic, lowering the temperature by $1keV$ in the hohlraum gas fill. It is important to note that the overall impact of the self-generated magnetic field is still to increase the hohlraum temperature, as will be shown later; the Larmor correction simply reduces the effect of MHD. It is not just the heat-flow that is being enhanced by the Larmor correction; the Nernst effect is also enhanced, which then acts to demagnetize the plasma.


 
\begin{figure}
 	\centering
 	\centering
 	\includegraphics[scale=0.7]{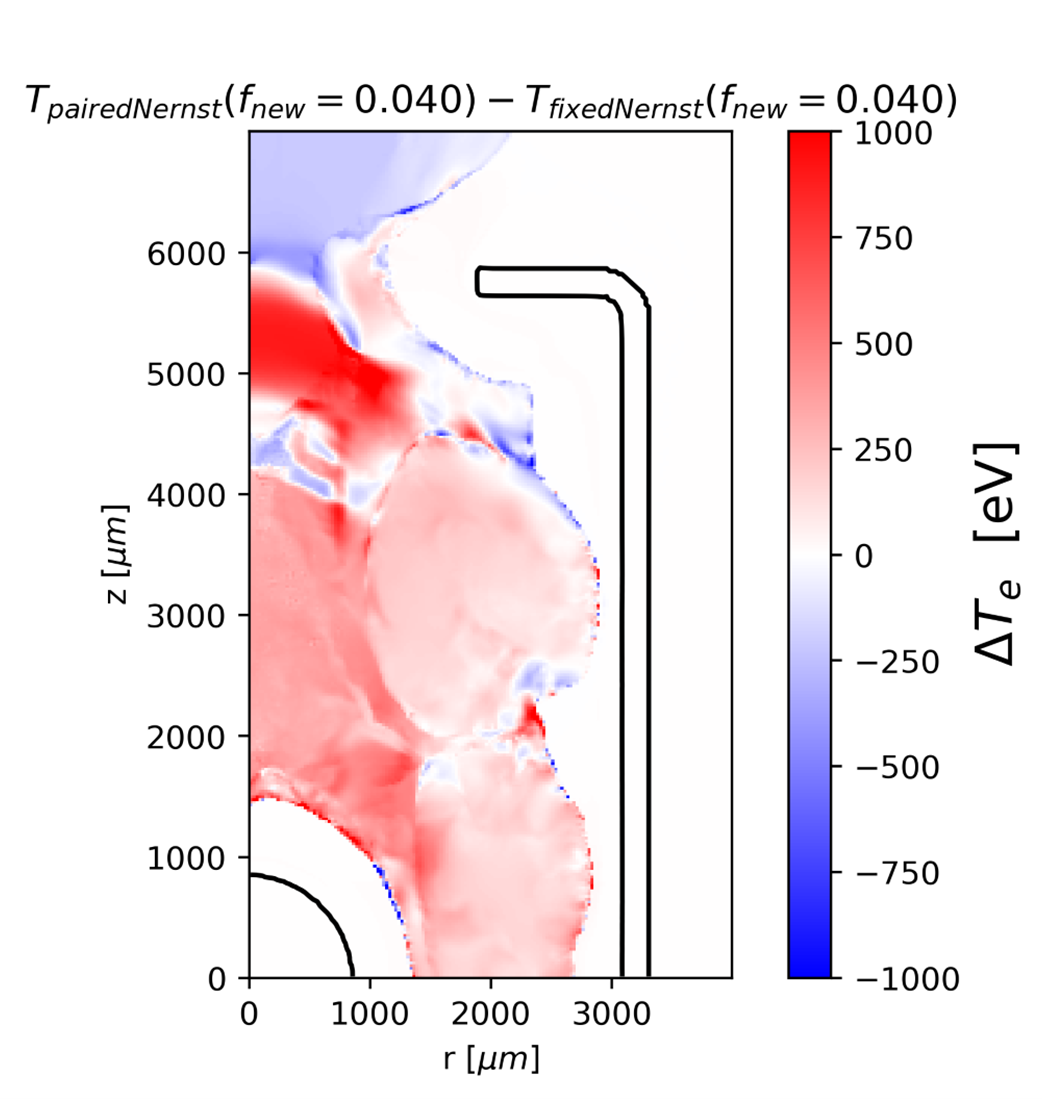}\caption{ \label{fig:hohlraum_nernst} Change in electron temperature in a hohlraum simulation due to changing from a fixed Nernst flux limiter (equation \ref{eq:vN_fixed}) to a paired limiter (equation \ref{eq:vN_paired}). The paired limiter is supported by VFP data and should be adopted for code implementation. }	
\end{figure}

The importance of implementing the Nernst flux limiter correctly is demonstrated in figure \ref{fig:hohlraum_nernst}, which shows the difference in temperature when moving from a fixed Nernst flux limiter to the recommended paired Nernst limiter. Throughout most of the hohlraum the temperature is higher, up to 1keV in excess. The paired limiter suppresses Nernst more, which results in higher magnetizations. Figure \ref{fig:Harmonic_nernst} also showed that the difference in flux limiter implementation is more important for low Z plasmas; this is borne out here through the highest temperature increase being seen in the helium gas fill.

	

 \begin{figure}
 	\centering
 	\centering
 	\includegraphics[scale=0.7]{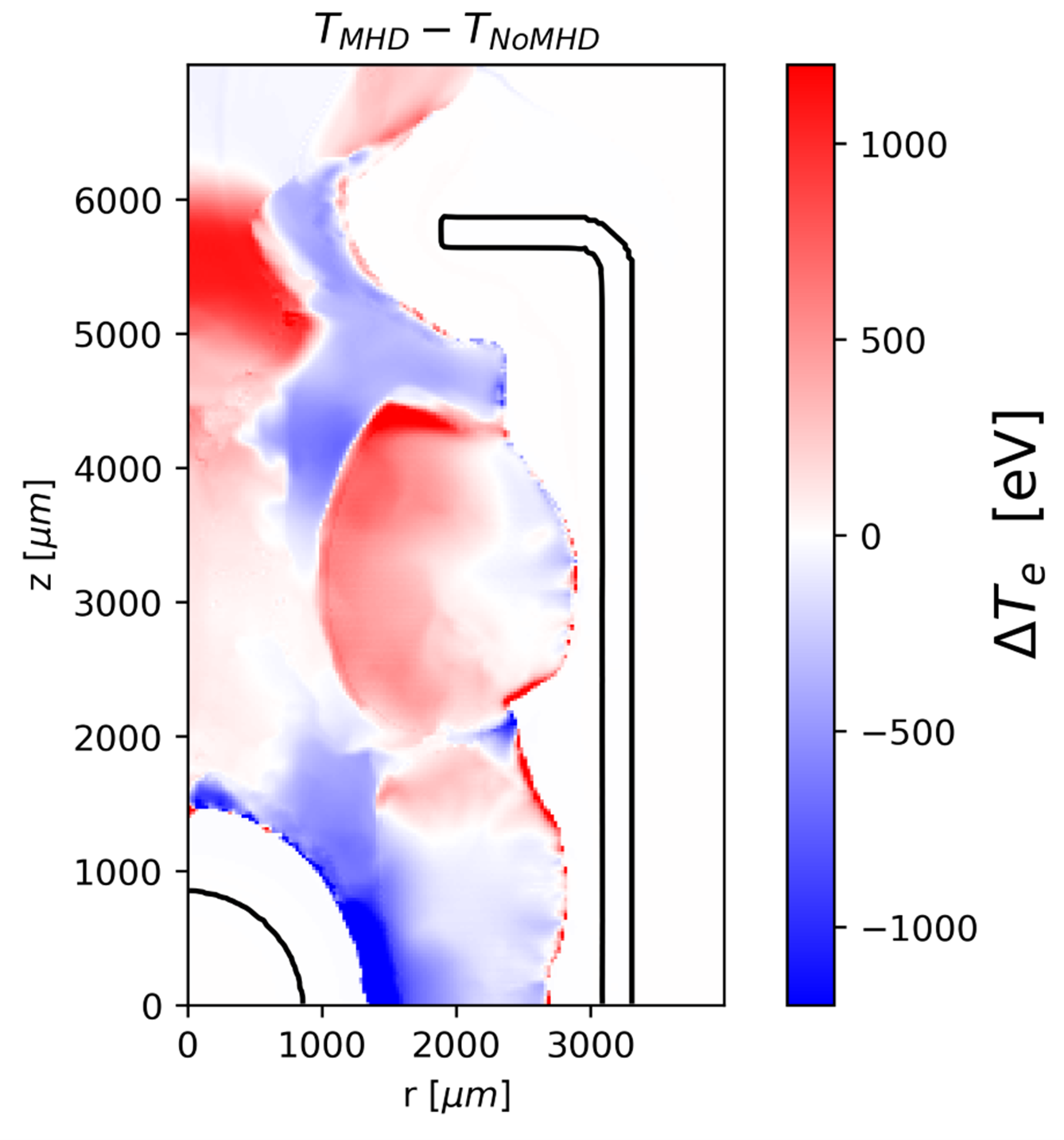}\caption{ \label{fig:hohlraum_MHD} The overall change in electron temperature calculated in a hohlraum due to self-generated magnetic fields using flux limiter methods that agree well with the VFP data (i.e. using equations \ref{eq:q_larm}, \ref{eq:vN_paired}, \ref{eq:righi_larm} and \ref{eq:cross}). m=1.0 has been used. }	
 \end{figure}

The total impact of self-generated magnetic fields on the hohlraum electron temperature is displayed in figure \ref{fig:hohlraum_MHD}. The flux limiters used are from equations \ref{eq:q_larm}, \ref{eq:vN_paired}, \ref{eq:righi_larm} and \ref{eq:cross}, which are the forms recommended here. The value of $m$ has been chosen as $1.0$, with relatively small changes to the temperature ($<200eV$) observed when moving to $m=1.1$.

The self-generated magnetic fields suppress heat-flow throughout much of the hohlraum, with temperatures increasing by more than $1keV$ at the copper bubble edge and at the LEH. 

\section{Conclusion}

Reduced models for electron transport in a plasma that has long mean free paths relative to temperature length-scales have been compared against kinetic Vlasov-Fokker-Planck simulations. This paper recommends using equations \ref{eq:q_harmonic_lambdad1},\ref{eq:q_larm}, \ref{eq:vN_paired}, \ref{eq:righi_larm} and \ref{eq:cross} to flux limit the parallel heat-flow, perpendicular heat-flow, Nernst effect, Righi-Leduc heat-flow and cross-gradient-Nernst effect, respectively. 

Improvements to the heat-flow flux limiter apply to both magnetized and unmagnetized simulations. A harmonic form (rather than hard cut-off) more closely follows the VFP data. The flux limiter value $f_{new}$ has been shown to vary with plasma ionization, although this variation is less stark when the non-locality length-scale $\lambda_d$ is used in the place of electron ion mean free path $\lambda_{ei}$, i.e. when electron-electron collisions are taken into account. 

While the new unmagnetized flux limiter can be tuned to give a similar plasma temperature as the traditional flux limiter in a specific location, spatial differences emerge. For example, hohlraum simulations were used here to demonstrate how the new flux limiter restricts heat-flow in low-Z plasmas more severely, while allowing more heat-flow at high-Z.

When heat-flow magnetization is important, the non-locality length-scale should be modified to incorporate finite Larmor radius effects (equation \ref{eq:lamb_b}). This results in the kinetic multiplier on the heat-flow being closer to unity with a magnetic field present and is most significant at moderate magnetization.

Nernst advection of magnetic fields down temperature gradients must also be flux-limited. The best approach is to limit Nernst by the same fraction that the heat-flow is limited, i.e the ratio of heat-flow velocity to the Nernst velocity remains approximately constant even when in the kinetic regime. Again, the importance of this approach is demonstrated using hohlraum simulations, with temperatures in the low density gas fill increasing by 1keV when the paired limiter is adopted.

While excellent agreement was reached between the theoretical flux limiters and kinetic data, this does not make future VFP simulations redundant. As the magnitude of the temperature perturbation is changed, the flux limiter values needed to reproduce the data also vary. In addition, the reduced kinetic models do not retain any time-dependent behavior; the model assumes the heat-flow has reached some steady state value instantaneously. While there are significant limitations to using radiation-hydrodynamics or extended-magnetohydrodynamics frameworks to model kinetic systems (and tuning a flux limiter will never reproduce all aspects of an experiment), the suggested flux limiter improvements should nonetheless make reduced kinetic models more faithful to their more rigorous counterparts.  

More than anything, the conclusion of this paper is that great care should be taken when utilizing an extended-MHD model in a laser-heated plasma. Seemingly small changes to the implementation of different electron transport terms can result in drastically differing plasma conditions. For complex integrated systems, such as hohlraum experiments on the National Ignition Facility, this work motivates simplified experiments where models can be verified against clear data.

\section*{Acknowledgements}
This work was performed under the auspices of the U.S. Department of Energy by Lawrence Livermore National Laboratory under Contract DE-AC52-07NA27344. 

This document was prepared as an account of work sponsored by an agency of the United States government. Neither the United States government nor Lawrence Livermore National Security, LLC, nor any of their employees makes any warranty, expressed or implied, or assumes any legal liability or responsibility for the accuracy, completeness, or usefulness of any information, apparatus, product, or process disclosed, or represents that its use would not infringe privately owned rights. Reference herein to any specific commercial product, process, or service by trade name, trademark, manufacturer, or otherwise does not necessarily constitute or imply its endorsement, recommendation, or favoring by the United States government or Lawrence Livermore National Security, LLC. The views and opinions of authors expressed herein do not necessarily state or reflect those of the United States government or Lawrence Livermore National Security, LLC, and shall not be used for advertising or product endorsement purposes.

\section{Appendix: Code Implementation \label{sec:Code}}

This section contains additional details helpful to those implementing these new flux limiter forms into extended-MHD codes. 

First, the temperature length-scale. For codes without magnetic fields the length-scale is simply $L_T = T_e/|\nabla T_e |$. Once a magnetic field is introduced it is important to distinguish the components parallel and perpendicular to the field. The simplest component to calculate is:
\begin{equation}
L_{T\parallel} = \frac{T_e}{|\underline{\hat{b}}\cdot\nabla T_e|} 
\end{equation}
Where $\underline{\hat{b}}$ is the magnetic field unit vector. From this, the perpendicular length-scale can be calculated:
\begin{equation}
L_{T\bot} = \sqrt{\frac{L_T ^2 L_{T\parallel}^2}{L_{T\parallel}^2 - L_T ^2}}
\end{equation}

For calculating the Righi-Leduc flux limiter in equation \ref{eq:righi_larm}, the maximum value of the Righi-Leduc coefficient for a given plasma ionization is needed. To enable fast calculation in the Gorgon MHD code, the following fit was found to be accurate with $<1\%$ error:

\begin{equation}
\kappa_{\wedge,max} = \frac{0.1183 + 1.894 Z +0.3343 Z^2}{1+0.5985 Z + 0.06385 Z^2} \label{eq:kappa_max}
\end{equation}

Note that for each heat-flow component relative to the magnetic field, the total heat-flow relative to the simulation grid should be calculated. For example, the unmagnetized isotropic heat-flow is typically calculated at cell faces. For a Cartesian system, the x-directed unmagnetized heat-flow should be flux limited using the total unmagnetized heat-flow at the x-faces. If this approach isn't taken, heat-flows that aren't directed along the cardinal directions are less flux limited.

	\section*{Bibliography}
	
	\bibliographystyle{unsrt}
	\bibliography{library}

\end{document}